\documentclass[prl,twocolumn,floatfix,notitlepage,superscriptaddress,longtable,nofootinbib]{revtex4-2}
\usepackage{amsmath}
\usepackage{lipsum} 
\usepackage{verbatim}
\usepackage{epsfig}
\usepackage{subfigure}
\usepackage{graphicx}
\usepackage{enumitem}
\usepackage{amsfonts}
\usepackage[final]{showlabels}
\usepackage{enumitem}
\usepackage[figuresright]{rotating}
\usepackage{amssymb}
\usepackage{amsmath}
\usepackage{psfrag}
\usepackage{mathtools}
\usepackage[export]{adjustbox}
\usepackage{bm}
\usepackage[colorlinks,linkcolor=blue,anchorcolor=blue,citecolor=blue,urlcolor=blue]{hyperref}
\usepackage[version=4]{mhchem}
\usepackage[svgnames]{xcolor}

\def\be{\begin{equation}} \def\ee{\end{equation}}
\def\bea{\begin{eqnarray}} \def\eea{\end{eqnarray}}

\def\bpm{\begin{pmatrix}} \def\epm{\end{pmatrix}}

\DeclareMathOperator{\Tr}{Tr}

\DeclareMathOperator{\diag}{diag}

\definecolor{Qicolor}{RGB}{3, 136, 252}

\makeatletter
\newcommand*{\balancecolsandclearpage}{%
  \close@column@grid
  \clearpage
}
\makeatother

\begin{document}

\title{Solvable model of quantum Darwinism-encoding transitions}

\author{Beno\^it Fert\'e}
\affiliation{Universit\'e Paris-Saclay, CNRS, LPTMS, 91405, Orsay, France}

\author{Xiangyu Cao}
\affiliation{Laboratoire de Physique de l'\'Ecole normale sup\'erieure, ENS, Universit\'e PSL, CNRS, Sorbonne Universit\'e, Universit\'e Paris Cit\'e, F-75005 Paris, France}

\date{\today}

\begin{abstract}
We propose a solvable model of Quantum Darwinism to encoding transitions---abrupt changes in how quantum information spreads in a many-body system under unitary dynamics. We consider a random Clifford circuit on an expanding tree, whose input qubit is entangled with a reference. The model has a Quantum Darwinism phase, where one classical bit of information about the reference can be retrieved from an arbitrarily small fraction of the output qubits, and an encoding phase where such retrieval is impossible. The two phases are separated by a mixed phase and two continuous transitions. We compare the exact result to a two-replica calculation. The latter yields a similar ``annealed'' phase diagram, which applies also to a model with Haar random unitaries. We relate our approach to measurement induced phase transitions (MIPTs), by solving a modified model where an environment eavesdrops on an encoding system. It has a sharp MIPT only with full access to the environment. 
\end{abstract}
\maketitle

\noindent\textbf{Introduction} 
A pillar of modern quantum statistical mechanics ~\cite{deutsch,srednicki,rigol-review} is the idea that unitary dynamics in a many-body system generically scrambles local quantum information. Eventually, it becomes highly nonlocal and impossible to retrieve, unless the observer has access to more than half of the system: the information has been {encoded}~\cite{nielsen_chuang_2010,preskilllecture,schumacher,schumacher-QEC}. Information scrambling and encoding have far-reaching consequences, for example on the quantum physics of black holes~\cite{Sekino_2008,haydenpreskill,shenkerstanford,swingle-review,brown19-teleport,schuster22-tele}. 

Meanwhile, a basic premise of Quantum Darwinism (QD)~\cite{blume-kohout-zurek,ollivier-poulin,zurek-QD,paternostro18-darwin,unden19-darwin-exp,zurek-review} is that a macroscopic environment, e.g., a measurement apparatus, \textit{duplicates} some classical information. Hence, the latter becomes retrievable in multiple small fractions of the environment. It is important to view the environment itself as a many-body quantum system. Indeed, the theory of QD aims to {deduce} the properties of the classical world from the core principles of quantum physics. According to QD, the duplication of information underlies the emergence of classical \textit{objectivity}~\cite{Korbicz-prl,le-olaya-pra,le-prl19,Korbicz-rev}: being objective is being known to many.

\begin{figure}
    \centering
    \includegraphics[width=\columnwidth]{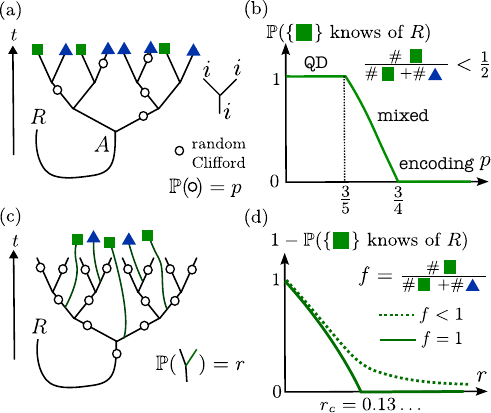}
    \caption{(a) Model for Quantum Darwinism-encoding transitions on an expanding tree with $t=3$ generations. (b) Information on $R$ is accessible to a small subsystem (squares) in the Quantum Darwinism (QD) phase, and inaccessible in the encoding phase. In the mixed phase, the information is accessible in a fraction of random realizations. (c) A tree model of an environment eavesdropping on an encoding dynamics. (d) A transition is only possible with full access to the environment $f=1$.} %
    \label{fig:fig1}
\end{figure}
Quantum Darwinism and encoding are distinct ways of many-body quantum information spreading. Both behaviors emerge from the microscopic laws of quantum mechanics, just like both ferro- and para-magnetism can emerge from the Ising model. Ferro- and para-magnetism are distinct phases of matter, separated by a continuous phase transition. Can we view QD and encoding as stable phases of quantum information, and are they separated by some transition~\cite{riedel2012,campbell}? In this Letter, we propose a solvable model of sharp phase transitions from QD to encoding. Our model is a random Clifford unitary circuit on an expanding tree, whose root forms a maximally entangled pair with a reference qubit [Fig.~\ref{fig:fig1}-(a)]. It has one parameter, analogue of the temperature in the Ising model. We then ask whether it is possible to retrieve information about the reference bit from a small fraction $f<1/2$ of the tree's leaves (output qubits). We determine exactly the model's phase diagram [Fig.~\ref{fig:fig1}-(b)]. It has a stable QD (encoding, resp.) phase, where one may (may not, resp.) extract a classical bit of information about the reference bit. Unlike the Ising model, the encoding and QD phases are separated by an intermediate mixed phase and two continuous transitions.

Another inspiration for this work is the measurement-induced phase transitions (MIPT) \cite{skinner19,li-fisher,amoschan,vasseur-ludwig,cao-deluca,schomerus19,li-fisher2,choi-altman-prl,bao-altman,vasseur-ludwig,gullans-huse-prx,schiro,MIPTrev}, which are also ``quantum information transitions''. In the standard setup, a generic many-body unitary evolution is continually interrupted by local measurements. By tuning the measurement rate, one obtains a transition between a phase with volume-law entanglement entropy and one with area law. The MIPTs concern entanglement properties of random states drawn from the Born rule, and are delicate to study and observe~\cite{li2022cross,khemani-post,MIPTnature}. Here, we consider a ``Darwinian'' MIPT setup, see Fig.~\ref{fig:fig1}-(c,d). We amend our model in the encoding phase with eavesdropping qubits~\cite{girolami}, and ask whether they can extract a classical bit of information about the reference~\cite{bao-altman,gullans-huse-prl,Vijay-Vishwanath,li-fisher-QEC,vijay-encoding}. We show that a sharp transition occurs at a critical rate of eavesdropping, if and only if one has access to all the eavesdropping bits. 

\noindent\textbf{Model for QD-encoding transition} 
Consider a maximally entangled pair 
$  ( |0\rangle_R \vert 0 \rangle_A + |1\rangle_R \vert 1 \rangle_A)  / \sqrt{2}$
between a reference qubit $R$ that will be kept intact, and the qubit $A$ that will be the root of an expanding binary tree unitary circuit, see Fig.~\ref{fig:fig1}. The edges of the tree represent the world lines of the qubits constituting a growing system~\cite{nahum21,feng2022measurement}. At each branching, we recruit a new qubit with state $|0\rangle$, and apply a \texttt{CNOT} gate to it and the input qubit: 
\begin{equation}\label{eq:CNOT}
    \includegraphics[valign=c,scale=1.4]{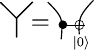} \,.
\end{equation}
Equivalently, the branching acts on the input qubit as an isometry $ \sum_{i=0,1}  |ii \rangle \langle i \vert $. 
In addition, we apply a random one-body Clifford unitary (drawn uniformly) to each edge of the tree with probability $p$, which is the parameter that interpolates between the QD ($p=0$) and encoding limits ($p=1$). After $t$ time steps, there are $N = 2^t$ output qubits, from which we draw the subsystem $F$ randomly: each output qubit belongs to $F$ with probability $f$. We denote by $U$ the resulting unitary from $A$ and $N-1$ recruits to the $N$ output qubits. By construction, $U$ is a Clifford unitary, which can be efficiently simulated~\cite{gottesman1998heisenberg,aaronson-gottesman}. Here, we can analyze the knowledge of $F$ on $R$ analytically~\cite{supp}.

For this, we recall the defining property of a Clifford unitary: it transforms any Pauli operator to a \textit{single} product of Pauli's, known as a Pauli string. For example, a one-body Clifford unitary permutes $X,Y$ and $Z$, and choosing a random one-body Clifford amounts to picking one among the 6 permutations (here and below, a Pauli string will be always considered modulo a phase $\pm1,\pm i$). Now, let us fix a realization of our model, and consider a Pauli string $P$ acting on the subsystem $F$. By definition, our Clifford unitary $U$ will pull it back to $Q = U^{\dagger}P U$, a Pauli string acting on $A$ and the $N-1$ recruits. We then contract it with the recruit states $(|0\rangle \langle 0|)^{\otimes N-1}$ to obtain a Pauli operator $O_A$ acting on $A$. There are two possibilities: (1) if $Q$ contains an $X$ or $Y$ acting on some recruit bit, $O_A$ vanishes. (2) Otherwise, $O_A \in \{I,Z, X, Y\}$ is identity or a Pauli. 

Repeating this for all Pauli strings acting on $F$, we construct a set $\mathbf{s} \subset \{I, X, Y, Z\}$ of all the nonzero operators $O_A$ thus obtained. It is not hard to see that $\mathbf{s}$ is a subgroup of  $\{I,X,Y,Z\}$ (modulo phase), i.e., $\mathbf{s}$ must equal one of these:
\begin{align}
& \mathbf{n} = \{I\}, \mathbf{z} = \{ I, Z \},  \mathbf{x} = \{ I, X \} ,  \mathbf{y} = \{ I, Y \}, \nonumber  \\ 
& \mathbf{a} = \{ I, Z, X, Y\} . 
\end{align} 
Since $RA$ is initially a maximally entangled pair, $\mathbf{s}$ tells us exactly what information about $R$ is accessible from $F$. If $\mathbf{s} = \mathbf{n}$, $F$ is uncorrelated with $R$. If $\mathbf{s} = \mathbf{z},  \mathbf{x}$ or $\mathbf{y}$, $F$ contains one classical bit of information on $R$: some Pauli string $O_F$ on $F$ is perfectly correlated with $O_R = Z$, $X$ or $Y$ on $R$. {More precisely, $O_F O_R$ is a {stabilizer} of the output state $\Psi_t$: $O_F O_R |\Psi_t \rangle  = \pm |\Psi_t \rangle $}. If $ \mathbf{s} = \mathbf{a}$, one may distill from $F$ a qubit maximally entangled with $R$~\cite{supp}. 

\begin{figure}
    \centering
    \includegraphics[width=\columnwidth]{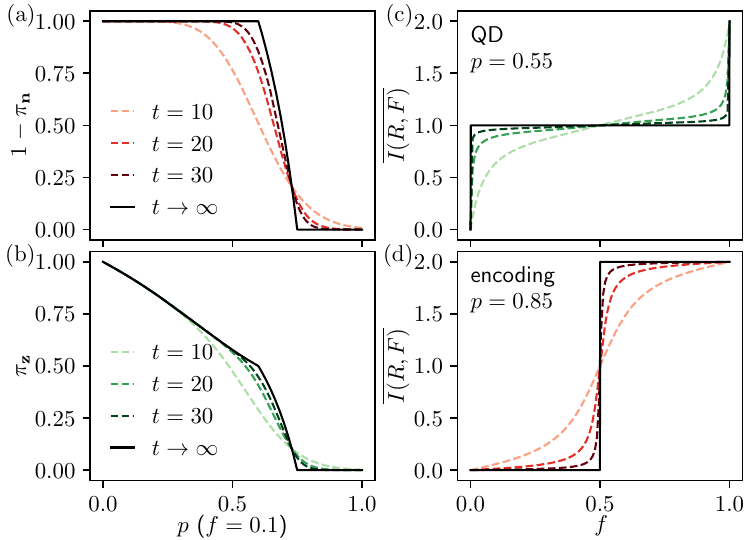}
    \caption{(a,b) $p$-dependence of the order parameters $1-\pi_{\mathbf{n}}$ (identical to Fig.~\ref{fig:fig1}-b) and $\pi_{\mathbf{z}}$. (c,d) Averaged mutual information $\overline{I(R, F)}$ as a function of the size fraction $f$ in the QD and encoding phase (resp.). { The average is over the random Clifford unitary and the random subset $F$}.  The finite $t$ data are from numerical iteration of the backward recursion, and the $t=\infty$ curves are the exact prediction~\cite{supp}. }
    \label{fig:num} 
\end{figure}

\noindent\textbf{Phase diagram} 
The ``order parameter'' of our model is thus the probability distribution of $\mathbf{s}$:
\begin{equation}\label{eq:pi}
     {\pi} := (\pi_{\mathbf{n}},  \pi_{\mathbf{z}}, \pi_{\mathbf{x}},  \pi_{\mathbf{y}},  \pi_{\mathbf{a}}) \,,
\end{equation}
where $\pi_{\mathbf{n}}$ is the probability that $\mathbf{s} =\mathbf{n}$, and so on. We can compute ${\pi}$ of a tree with $t$ generations from one with $(t-1)$ using a ``backward recursion'' relation.  The phase diagram of the model is determined by iterating this relation and analyzing the $t\to\infty$ limit of ${\pi}$ as a function of $p$ (and $f$)~\cite{supp}. As a result, we find three phases, see Fig.~\ref{fig:fig1}-(b) for a sketch and Fig.~\ref{fig:num} for plots. When $p < 3/5$, we have a Quantum Darwinism (QD) phase, where for any $f \in (0,1)$, we have $\pi_{\mathbf{a}}\to0, \pi_{\mathbf{n}} \to 0$, and 
    \begin{equation}\label{eq:QD-sol}
        \pi_{\mathbf{z}} \to \frac{3-6p +\sqrt{24 (p-1) p+9}}{6 - 6 p}  \,, \pi_{x,y} \to \frac{1-\pi_{\mathbf{z}}}2 \,.
    \end{equation}  
($\pi_{{\mathbf{z}}} \to 1$ as $p\to 0$.) When $p > 3/4$, we have a \textit{encoding} phase, where $\pi_{\mathbf{n}} \to 1$ if $f < 1/2$ and $\pi_{\mathbf{a}} \to 1$ if $f > 1/2$. 
Finally, when $3/5<p<3/4$, we have a mixed phase. For any $f<1/2$, we have $\pi_{\mathbf{a}} \to 0$ while
\begin{equation} \label{eq:mixedstate}
    (\pi_{\mathbf{n}}, \pi_{\mathbf{z}}, \pi_{\mathbf{x}}, \pi_{\mathbf{y}}) \stackrel{f<\frac12}\to (1- u, \frac{u}2, \frac{u}4 , \frac{u}4),\, u = \frac{6-8p}{3-3p} \,.
\end{equation}
Here $u$ is probability that we can retrieve one classical bit from the subsystem $F$, and it decreases from $1$ to $0$ as $p$ varies from $3/5$ to $3/4$. The solution for $f>1/2$ is obtained from \eqref{eq:mixedstate} by swapping $ \pi_{\mathbf{n}}$ and $\pi_{\mathbf{a}}$. 
 
The existence of the two transitions, { at $p = 3/5$ and $p = 3/4$ respectively, where $\pi$ is non-analytical,} can be associated to the breaking/restoration of two symmetries of the model. First, a $\mathbb{Z}_2$ symmetry acts by exchanging $\pi_{\mathbf{n}} \leftrightarrow \pi_{\mathbf{a}}$, or swapping the subsystem $F$ and its complement (without $R$)~\cite{tianci}. This symmetry is preserved by the circuit dynamics, weakly broken by the ``boundary condition'' (the choice of $F$), and restored only in the QD phase. Second, a $\mathcal{S}_3$ symmetry acts by permuting $\mathbf{x}, \mathbf{y}, \mathbf{z} $ (while leaving $\mathbf{n}$ and $\mathbf{a}$ invariant). This symmetry is preserved by the random one-body Clifford unitary, broken by the branching~\eqref{eq:CNOT}, and restored only in the encoding phase. The mixed phase breaks both symmetries. We numerically explored a few other Clifford variants of our model, and found the above two-stage scenario to be rather general~\cite{inprep}. 

\noindent\textbf{Mutual information and discord}
It is useful to consider the mutual information between $F$ and $R$, defined as $I(R,F) = H(R) + H(F) - H(RF)$, where $H(X)= - \mathrm{Tr}[\rho_X \log_2 \rho_X]$ is the von Neumann entropy. In our model, it is not hard to see that $I(R,F) = \log_2 |\mathbf{s}| $ is the dimension of $\mathbf{s}$ as a vector space over $\mathbb{Z}_2$~\cite{supp}. So, in the QD phase, 
\begin{equation} \label{eq:darwinism-plateau}
     I(R,F)  \to  1  \quad (0<f<1)  \quad  \text{(QD)} \,,
\end{equation}
with probability one [Fig.~\ref{fig:num}-(c)]. The independence of  $I$ on the fraction size $f$, sometimes called the ``objectivity plateau'', is a hallmark of QD~\cite{zurek-QD}. Meanwhile, in the encoding phase, 
\begin{equation} \label{eq:encoding}
     I(R,F) \to  \begin{dcases} 
    0 & f < 1/2\\
    2  & f > 1/2
    \end{dcases} \quad  \text{(encoding)}
\end{equation}
with probability one [Fig.~\ref{fig:num}-(d)], as expected from the Page curve~\cite{page}. In the mixed phase, we may wonder how the $I$-$f$ curve looks like in a \textit{single} realization (with large $t$), where we increase $f$ by gradually adding random qubits into $F$. To address this question, we computed the joint distribution of $(\mathbf{s}, \mathbf{t})$ corresponding to two random subsystems $F \subset G$, and a same unitary $U$~\cite{supp}. As a result, we found that a single-realization $I$-$f$ curve is exactly the QD one \eqref{eq:darwinism-plateau} with probability $u$ defined in \eqref{eq:mixedstate}, and exactly the encoding curve \eqref{eq:encoding} with probability $1-u$. In other words, the intermediate-phase ensemble is a mixture of QD and encoding realizations, both occurring with nonzero probability in the $t\to\infty$ limit. 

In general, the mutual information between $F$ and $R$ does not correspond exactly to the amount of information that one can learn about $R$ by observing $F$~\cite{discord-zurek,Henderson_2001}. The discrepancy is known as ``quantum discord''. Here, the discord vanishes whenever $I(R,F) = 1$, given the knowledge of the unitary circuit: we can construct the observable on $F$ which reveals the classical bit of information on $F$. Moreover, we can show that in the QD phase, one may still retrieve a bit of information from $R$ even with access to only  the $Z$ operators on $F$.

\begin{figure}
    \centering
    \includegraphics[width=1\columnwidth]{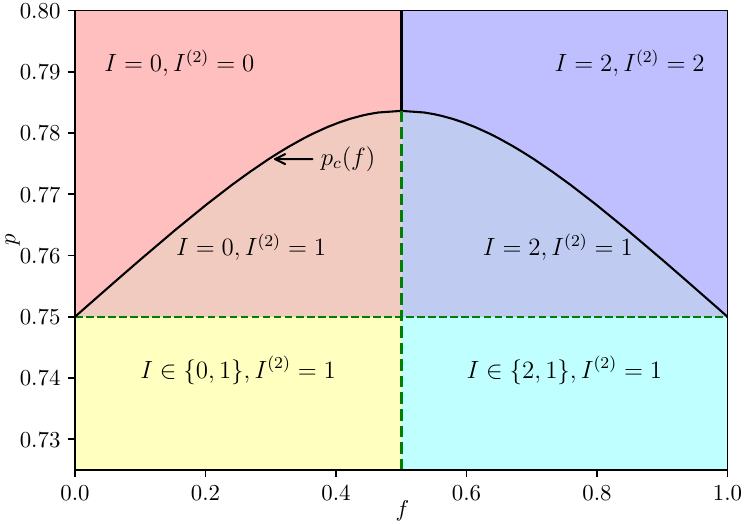}
    \caption{Comparing the annealed mutual information $  I^{(2)}(F, R) $~\eqref{eq:I2} with the genuine one $I(F,R)$. They disagree in the mixed phase ($3/5 < p< 3/4$) and in part of the encoding phase where $3/4 < p < p_c(f)$. $p_c(f)$ (solid curves) is determined numerically using the recursion relation for $I^{(2)}$~\cite{supp}. }
    \label{fig:I2}
\end{figure}
\noindent\textbf{Two-replica analysis} 
A valuable tool to compute quantum information quantities is the ``replica trick''~\cite{wilczek,CalabreseCardy_2004,tonni-negativity,lashkari,ruggiero-trace-dis,kudler-renyiMI}. Yet, results of replica calculations can be subtle to interpret, especially if one is not able to take the appropriate replica number limit. Here, we perform a two-replica analysis of our model, and compare the result with the exact phase diagram. 

In the replica approach, the accessible quantity is the ``annealed'' mutual information 
\begin{equation}\label{eq:I2def}
    I^{(2)}(F, R) := \log_2   \mathrm{Tr}\left[\overline{\rho_{FR}^2}\right] - \log_2 \mathrm{Tr}\left[\overline{\rho_{F}^2}\right] + 1 \,,
\end{equation} 
where $\overline{[\dots]}$ denotes an average over $U$ and $F$. Note that $I^{(2)}$ would equal to the average von Neumann mutual information if $\mathrm{Tr}[\overline{\rho_{X}^2}]$ {were} equal to $ 2^{-\overline{H(X)}}$ (which is wrong!). The annealed mutual information can be computed by random unitary circuit techniques~\cite{nahum-entanglement,nahum-otoc,tianci,MIPTrev,colin}; indeed, since the Clifford group is a 2-design~\cite{livine-2design}, $ I^{(2)}(F, R)$ will not change if we replace a random one-body Clifford unitary with a Haar-random one in $U(2)$. We find~\cite{supp}:
\begin{equation}\label{eq:I2}
    I^{(2)}(F, R)  \to \begin{dcases}
          0 & f < 1/2, p > p_c(f) \\
        2 & f > 1/2, p > p_c(f) \\
       1 & p < p_c(f)
    \end{dcases} \,.
\end{equation}
Here $p_c(f) = p_c(1-f)$ is a threshold function that increases from  $p_c(0) = 3/4$ to $p_c(1/2) = \frac{3}{7} \left(2 \sqrt{2}-1\right) = 0.783\dots$, see Fig.~\ref{fig:I2}.

The ``annealed phase diagram'' of $I^{(2)}$ is similar to the exact one, with however differences: $I^{(2)}(F, R) = 1$ in both QD and mixed phases, as well as a small part of the encoding phase. So, the  annealed phase diagram is biased towards QD, which we qualitatively explain as follows. Both purity averages in \eqref{eq:I2def} are dominated by realizations with small entanglement entropy in $F$. Now, QD states tend to have low entanglement; indeed, the ``perfect'' QD-state (produced at ${p=0}$) is the GHZ state~\cite{greenberger1989}, 
$$
    | \text{GHZ} \rangle = \frac1{\sqrt{2}}( |0_R \underbrace{ 0 \dots 0}_F  0 \dots 0 \rangle +  |1_R \underbrace{ 1 \dots 1}_F  1 \dots 1  \rangle ) \,.
$$
It has one bit of entanglement entropy for any bipartition. In comparison, an encoding state has a volume law entropy. Hence, in both QD and mixed phases, QD realizations will dominate $I^{(2)}$, which fails to distinguish them. In the encoding phase, a QD realization occurs with an exponentially small (in $t$) probability, yet its $\Tr[\rho_F^2]$ and $\Tr[\rho_{FR}^2]$ can be exponentially large compared to the typical encoding states. Hence, rare QD states in the encoding phase can dominate the annealed mutual information. 

\noindent\textbf{Relating to MIPT}
 The QD-encoding transitions (QDETs) differ from the measurement induced ones (MIPTs) in two ways. First, MIPTs result from the competition between a scrambling system and its environment (the measurement apparatus). { Meanwhile, QDETs take place within a \textit{structured} environment~\cite{Ryan-onion}. In the QD phase, the environment behaves as a macroscopic apparatus that ``measures'' the reference spin in some direction (the direction is $Z$ with probability $\pi_{\mathbf{z}}$, and so on), and broadcast the outcome. As we tune the apparatus into the encoding phase, it becomes dysfunctional and fails to broadcast any information on the reference system.} 
 
 Second, QDETs are about the information available in small environment fractions, while MIPTs are observable only with full access to the environment. 
 {To support this claim, we consider a variant of our model that mimics the MIPT setup}. We take the above model at $p=1$ (in the encoding phase), and let every qubit in the tree be subject to an eavesdropping event with probability $r$. The eavesdropping consists again as a branching~\eqref{eq:CNOT}, of which one output bit is then emitted to the ``environment'', see Fig.~\ref{fig:fig1}-(c). After $t$ generations, we have a system with $N = 2^t$ bits and an environment $E$ of average size $|E| = (2N-1)r$. 
 
 Then we ask: can we retrieve information on $R$ from a fraction $F$ of the \textit{environment}, with $|F| /  |E| = f$?  Moreover, we only allow access to $Z$ operators on $F$ (allowing access to all operators results in an entirely different phase diagram~\cite{inprep}). Then, the order parameter \eqref{eq:pi} obeys a modified recursion relation~\cite{supp}. In particular, ${\pi}_{\mathbf{a}} = 0$, and the probability of retrieving one classical bit equals $1 - \pi_{\mathbf{n}}$. We find that, when $f = 1$, there is a transition:
\begin{equation} \label{eq:MIPT}
    \pi_{\mathbf{n}} \stackrel{f=1} \to  \begin{dcases}
    \frac{4 r^2 - 8 r + 1}{1-r}  & r < r_c \\ 
    0 & r > r_c \,,
    \end{dcases}
\end{equation}
where $r_c= \frac{1}{2} \left(2-\sqrt{3}\right) \approx 0.134$. This transition is equivalent to the standard MIPT. Indeed, consider projectively measuring $Z$ on all the qubits of $F$. If $\mathbf{s} = \mathbf{n}$, the measurements reveal nothing about $R$, which remains entangled with unmeasured bits. Otherwise, if say $\mathbf{s} = \mathbf{x}$, the measurements will project the qubit $R$ to an eigenstate of $X$, disentangling it. Therefore,  $r > r_c$ is the area-law (purified) phase and $r < r_c$ the volume-law (encoded) phase~\cite{bao-altman,gullans-huse-prl,vijay-encoding,li-fisher-QEC}. Note that the transition exists \textit{only} at $f = 1$, where almost all the environment is accessible. For any $f < 1$, $ \pi_{\mathbf{n}}(t\to\infty)$ depends smoothly on $r$ and never vanishes. This is after all reasonable from the MIPT point of view: we need all the measurement outcomes to construct the quantum trajectory state. 

\noindent\textbf{Outlook}
We introduced a solvable model for Quantum Darwinism-encoding transitions (QDETs). They are a new type of quantum information phase transitions under unitary evolution, {where the different phases are characterized by whether information about the reference qubit is retrievable from small fractions of the environment.}
It will be interesting to identify QDETs in finite-dimensional ($d<\infty$) systems and characterize their universality classes; our tree model is equivalent to an all-to-all ($d=\infty$)  circuit, and has simple mean-field critical exponents~\cite{nahum2023renormalization}. In particular, it may be nontrivial to establish a QD phase in a $d<\infty$ geometry, which hinders the fast spread of information~\cite{LRbound,LRbound-LR,chen2023speed}; an expanding (de Sitter) geometry could be necessary.
Another important question concern QDETs in non-Clifford models~\cite{QD-QBM,girolami,riedel2012}, in particular, whether the mixed phase is generic. Indeed, the knowledge of $F$ on $R$ is in general not ``quantized'' as in a Clifford model. This will affect the nature of the order parameter, and make even the mean-field theory more involved~\cite{nahum21,feng2022measurement,ferte2023tree}.
Finally, encoding is proper to the quantum realm, and Quantum Darwinism is a theory of the emergence of the classical. Thus, we hope to shed light on the quantum-classical transition through the lens of dynamical critical phenomena.

\begin{acknowledgements}
We thank Andrea De Luca and the Anonymous Referee for helpful comments on the manuscript. X.C. acknowledges support from CNRS and ENS, and thanks LPTMS for hospitality. 
\end{acknowledgements}

\bibliography{ref}

\begin{thebibliography}{73}%
\makeatletter
\providecommand \@ifxundefined [1]{%
 \@ifx{#1\undefined}
}%
\providecommand \@ifnum [1]{%
 \ifnum #1\expandafter \@firstoftwo
 \else \expandafter \@secondoftwo
 \fi
}%
\providecommand \@ifx [1]{%
 \ifx #1\expandafter \@firstoftwo
 \else \expandafter \@secondoftwo
 \fi
}%
\providecommand \natexlab [1]{#1}%
\providecommand \enquote  [1]{``#1''}%
\providecommand \bibnamefont  [1]{#1}%
\providecommand \bibfnamefont [1]{#1}%
\providecommand \citenamefont [1]{#1}%
\providecommand \href@noop [0]{\@secondoftwo}%
\providecommand \href [0]{\begingroup \@sanitize@url \@href}%
\providecommand \@href[1]{\@@startlink{#1}\@@href}%
\providecommand \@@href[1]{\endgroup#1\@@endlink}%
\providecommand \@sanitize@url [0]{\catcode `\\12\catcode `\$12\catcode
  `\&12\catcode `\#12\catcode `\^12\catcode `\_12\catcode `\%12\relax}%
\providecommand \@@startlink[1]{}%
\providecommand \@@endlink[0]{}%
\providecommand \url  [0]{\begingroup\@sanitize@url \@url }%
\providecommand \@url [1]{\endgroup\@href {#1}{\urlprefix }}%
\providecommand \urlprefix  [0]{URL }%
\providecommand \Eprint [0]{\href }%
\providecommand \doibase [0]{https://doi.org/}%
\providecommand \selectlanguage [0]{\@gobble}%
\providecommand \bibinfo  [0]{\@secondoftwo}%
\providecommand \bibfield  [0]{\@secondoftwo}%
\providecommand \translation [1]{[#1]}%
\providecommand \BibitemOpen [0]{}%
\providecommand \bibitemStop [0]{}%
\providecommand \bibitemNoStop [0]{.\EOS\space}%
\providecommand \EOS [0]{\spacefactor3000\relax}%
\providecommand \BibitemShut  [1]{\csname bibitem#1\endcsname}%
\let\auto@bib@innerbib\@empty
\bibitem [{\citenamefont {Deutsch}(1991)}]{deutsch}%
  \BibitemOpen
  \bibfield  {author} {\bibinfo {author} {\bibfnamefont {J.~M.}\ \bibnamefont
  {Deutsch}},\ }\bibfield  {title} {\bibinfo {title} {Quantum statistical
  mechanics in a closed system},\ }\href
  {https://doi.org/10.1103/PhysRevA.43.2046} {\bibfield  {journal} {\bibinfo
  {journal} {Phys. Rev. A}\ }\textbf {\bibinfo {volume} {43}},\ \bibinfo
  {pages} {2046} (\bibinfo {year} {1991})}\BibitemShut {NoStop}%
\bibitem [{\citenamefont {Srednicki}(1994)}]{srednicki}%
  \BibitemOpen
  \bibfield  {author} {\bibinfo {author} {\bibfnamefont {M.}~\bibnamefont
  {Srednicki}},\ }\bibfield  {title} {\bibinfo {title} {Chaos and quantum
  thermalization},\ }\href {https://doi.org/10.1103/PhysRevE.50.888} {\bibfield
   {journal} {\bibinfo  {journal} {Phys. Rev. E}\ }\textbf {\bibinfo {volume}
  {50}},\ \bibinfo {pages} {888} (\bibinfo {year} {1994})}\BibitemShut
  {NoStop}%
\bibitem [{\citenamefont {D'Alessio}\ \emph {et~al.}(2016)\citenamefont
  {D'Alessio}, \citenamefont {Kafri}, \citenamefont {Polkovnikov},\ and\
  \citenamefont {Rigol}}]{rigol-review}%
  \BibitemOpen
  \bibfield  {author} {\bibinfo {author} {\bibfnamefont {L.}~\bibnamefont
  {D'Alessio}}, \bibinfo {author} {\bibfnamefont {Y.}~\bibnamefont {Kafri}},
  \bibinfo {author} {\bibfnamefont {A.}~\bibnamefont {Polkovnikov}},\ and\
  \bibinfo {author} {\bibfnamefont {M.}~\bibnamefont {Rigol}},\ }\bibfield
  {title} {\bibinfo {title} {From quantum chaos and eigenstate thermalization
  to statistical mechanics and thermodynamics},\ }\href
  {https://doi.org/10.1080/00018732.2016.1198134} {\bibfield  {journal}
  {\bibinfo  {journal} {Advances in Physics}\ }\textbf {\bibinfo {volume}
  {65}},\ \bibinfo {pages} {239} (\bibinfo {year} {2016})}\BibitemShut
  {NoStop}%
\bibitem [{\citenamefont {Nielsen}\ and\ \citenamefont
  {Chuang}(2010)}]{nielsen_chuang_2010}%
  \BibitemOpen
  \bibfield  {author} {\bibinfo {author} {\bibfnamefont {M.~A.}\ \bibnamefont
  {Nielsen}}\ and\ \bibinfo {author} {\bibfnamefont {I.~L.}\ \bibnamefont
  {Chuang}},\ }\href {https://doi.org/10.1017/CBO9780511976667} {\emph
  {\bibinfo {title} {Quantum Computation and Quantum Information: 10th
  Anniversary Edition}}}\ (\bibinfo  {publisher} {Cambridge University Press},\
  \bibinfo {year} {2010})\BibitemShut {NoStop}%
\bibitem [{\citenamefont {Preskill}(2023)}]{preskilllecture}%
  \BibitemOpen
  \bibfield  {author} {\bibinfo {author} {\bibfnamefont {J.}~\bibnamefont
  {Preskill}},\ }\href {http://theory.caltech.edu/~preskill/ph229/} {\bibinfo
  {title} {Lecture notes for ph219/cs219: Quantum information}} (\bibinfo
  {year} {2023})\BibitemShut {NoStop}%
\bibitem [{\citenamefont {Schumacher}(1995)}]{schumacher}%
  \BibitemOpen
  \bibfield  {author} {\bibinfo {author} {\bibfnamefont {B.}~\bibnamefont
  {Schumacher}},\ }\bibfield  {title} {\bibinfo {title} {Quantum coding},\
  }\href {https://doi.org/10.1103/PhysRevA.51.2738} {\bibfield  {journal}
  {\bibinfo  {journal} {Phys. Rev. A}\ }\textbf {\bibinfo {volume} {51}},\
  \bibinfo {pages} {2738} (\bibinfo {year} {1995})}\BibitemShut {NoStop}%
\bibitem [{\citenamefont {Schumacher}\ and\ \citenamefont
  {Westmoreland}(2002)}]{schumacher-QEC}%
  \BibitemOpen
  \bibfield  {author} {\bibinfo {author} {\bibfnamefont {B.}~\bibnamefont
  {Schumacher}}\ and\ \bibinfo {author} {\bibfnamefont {M.~D.}\ \bibnamefont
  {Westmoreland}},\ }\bibfield  {title} {\bibinfo {title} {Approximate quantum
  error correction},\ }\href {https://doi.org/10.1023/A:1019653202562}
  {\bibfield  {journal} {\bibinfo  {journal} {Quantum Information Processing}\
  }\textbf {\bibinfo {volume} {1}},\ \bibinfo {pages} {5} (\bibinfo {year}
  {2002})}\BibitemShut {NoStop}%
\bibitem [{\citenamefont {Sekino}\ and\ \citenamefont
  {Susskind}(2008)}]{Sekino_2008}%
  \BibitemOpen
  \bibfield  {author} {\bibinfo {author} {\bibfnamefont {Y.}~\bibnamefont
  {Sekino}}\ and\ \bibinfo {author} {\bibfnamefont {L.}~\bibnamefont
  {Susskind}},\ }\bibfield  {title} {\bibinfo {title} {Fast scramblers},\
  }\href {https://doi.org/10.1088/1126-6708/2008/10/065} {\bibfield  {journal}
  {\bibinfo  {journal} {Journal of High Energy Physics}\ }\textbf {\bibinfo
  {volume} {2008}},\ \bibinfo {pages} {065} (\bibinfo {year}
  {2008})}\BibitemShut {NoStop}%
\bibitem [{\citenamefont {Hayden}\ and\ \citenamefont
  {Preskill}(2007)}]{haydenpreskill}%
  \BibitemOpen
  \bibfield  {author} {\bibinfo {author} {\bibfnamefont {P.}~\bibnamefont
  {Hayden}}\ and\ \bibinfo {author} {\bibfnamefont {J.}~\bibnamefont
  {Preskill}},\ }\bibfield  {title} {\bibinfo {title} {Black holes as mirrors:
  quantum information in random subsystems},\ }\href
  {https://doi.org/10.1088/1126-6708/2007/09/120} {\bibfield  {journal}
  {\bibinfo  {journal} {Journal of High Energy Physics}\ }\textbf {\bibinfo
  {volume} {2007}},\ \bibinfo {pages} {120} (\bibinfo {year}
  {2007})}\BibitemShut {NoStop}%
\bibitem [{\citenamefont {Shenker}\ and\ \citenamefont
  {Stanford}(2014)}]{shenkerstanford}%
  \BibitemOpen
  \bibfield  {author} {\bibinfo {author} {\bibfnamefont {S.~H.}\ \bibnamefont
  {Shenker}}\ and\ \bibinfo {author} {\bibfnamefont {D.}~\bibnamefont
  {Stanford}},\ }\bibfield  {title} {\bibinfo {title} {Black holes and the
  butterfly effect},\ }\href {https://doi.org/10.1007/JHEP03(2014)067}
  {\bibfield  {journal} {\bibinfo  {journal} {Journal of High Energy Physics}\
  }\textbf {\bibinfo {volume} {2014}},\ \bibinfo {pages} {67} (\bibinfo {year}
  {2014})}\BibitemShut {NoStop}%
\bibitem [{\citenamefont {Swingle}(2018)}]{swingle-review}%
  \BibitemOpen
  \bibfield  {author} {\bibinfo {author} {\bibfnamefont {B.}~\bibnamefont
  {Swingle}},\ }\bibfield  {title} {\bibinfo {title} {Unscrambling the physics
  of out-of-time-order correlators},\ }\href
  {https://doi.org/10.1038/s41567-018-0295-5} {\bibfield  {journal} {\bibinfo
  {journal} {Nature Physics}\ }\textbf {\bibinfo {volume} {14}},\ \bibinfo
  {pages} {988} (\bibinfo {year} {2018})}\BibitemShut {NoStop}%
\bibitem [{\citenamefont {Brown}\ \emph {et~al.}(2023)\citenamefont {Brown},
  \citenamefont {Gharibyan}, \citenamefont {Leichenauer}, \citenamefont {Lin},
  \citenamefont {Nezami}, \citenamefont {Salton}, \citenamefont {Susskind},
  \citenamefont {Swingle},\ and\ \citenamefont {Walter}}]{brown19-teleport}%
  \BibitemOpen
  \bibfield  {author} {\bibinfo {author} {\bibfnamefont {A.~R.}\ \bibnamefont
  {Brown}}, \bibinfo {author} {\bibfnamefont {H.}~\bibnamefont {Gharibyan}},
  \bibinfo {author} {\bibfnamefont {S.}~\bibnamefont {Leichenauer}}, \bibinfo
  {author} {\bibfnamefont {H.~W.}\ \bibnamefont {Lin}}, \bibinfo {author}
  {\bibfnamefont {S.}~\bibnamefont {Nezami}}, \bibinfo {author} {\bibfnamefont
  {G.}~\bibnamefont {Salton}}, \bibinfo {author} {\bibfnamefont
  {L.}~\bibnamefont {Susskind}}, \bibinfo {author} {\bibfnamefont
  {B.}~\bibnamefont {Swingle}},\ and\ \bibinfo {author} {\bibfnamefont
  {M.}~\bibnamefont {Walter}},\ }\bibfield  {title} {\bibinfo {title} {Quantum
  gravity in the lab. i. teleportation by size and traversable wormholes},\
  }\href {https://doi.org/10.1103/PRXQuantum.4.010320} {\bibfield  {journal}
  {\bibinfo  {journal} {PRX Quantum}\ }\textbf {\bibinfo {volume} {4}},\
  \bibinfo {pages} {010320} (\bibinfo {year} {2023})}\BibitemShut {NoStop}%
\bibitem [{\citenamefont {Schuster}\ \emph {et~al.}(2022)\citenamefont
  {Schuster}, \citenamefont {Kobrin}, \citenamefont {Gao}, \citenamefont
  {Cong}, \citenamefont {Khabiboulline}, \citenamefont {Linke}, \citenamefont
  {Lukin}, \citenamefont {Monroe}, \citenamefont {Yoshida},\ and\ \citenamefont
  {Yao}}]{schuster22-tele}%
  \BibitemOpen
  \bibfield  {author} {\bibinfo {author} {\bibfnamefont {T.}~\bibnamefont
  {Schuster}}, \bibinfo {author} {\bibfnamefont {B.}~\bibnamefont {Kobrin}},
  \bibinfo {author} {\bibfnamefont {P.}~\bibnamefont {Gao}}, \bibinfo {author}
  {\bibfnamefont {I.}~\bibnamefont {Cong}}, \bibinfo {author} {\bibfnamefont
  {E.~T.}\ \bibnamefont {Khabiboulline}}, \bibinfo {author} {\bibfnamefont
  {N.~M.}\ \bibnamefont {Linke}}, \bibinfo {author} {\bibfnamefont {M.~D.}\
  \bibnamefont {Lukin}}, \bibinfo {author} {\bibfnamefont {C.}~\bibnamefont
  {Monroe}}, \bibinfo {author} {\bibfnamefont {B.}~\bibnamefont {Yoshida}},\
  and\ \bibinfo {author} {\bibfnamefont {N.~Y.}\ \bibnamefont {Yao}},\
  }\bibfield  {title} {\bibinfo {title} {Many-body quantum teleportation via
  operator spreading in the traversable wormhole protocol},\ }\href
  {https://doi.org/10.1103/PhysRevX.12.031013} {\bibfield  {journal} {\bibinfo
  {journal} {Phys. Rev. X}\ }\textbf {\bibinfo {volume} {12}},\ \bibinfo
  {pages} {031013} (\bibinfo {year} {2022})}\BibitemShut {NoStop}%
\bibitem [{\citenamefont {Blume-Kohout}\ and\ \citenamefont
  {Zurek}(2006)}]{blume-kohout-zurek}%
  \BibitemOpen
  \bibfield  {author} {\bibinfo {author} {\bibfnamefont {R.}~\bibnamefont
  {Blume-Kohout}}\ and\ \bibinfo {author} {\bibfnamefont {W.~H.}\ \bibnamefont
  {Zurek}},\ }\bibfield  {title} {\bibinfo {title} {Quantum darwinism:
  Entanglement, branches, and the emergent classicality of redundantly stored
  quantum information},\ }\href {https://doi.org/10.1103/PhysRevA.73.062310}
  {\bibfield  {journal} {\bibinfo  {journal} {Phys. Rev. A}\ }\textbf {\bibinfo
  {volume} {73}},\ \bibinfo {pages} {062310} (\bibinfo {year}
  {2006})}\BibitemShut {NoStop}%
\bibitem [{\citenamefont {Ollivier}\ \emph {et~al.}(2004)\citenamefont
  {Ollivier}, \citenamefont {Poulin},\ and\ \citenamefont
  {Zurek}}]{ollivier-poulin}%
  \BibitemOpen
  \bibfield  {author} {\bibinfo {author} {\bibfnamefont {H.}~\bibnamefont
  {Ollivier}}, \bibinfo {author} {\bibfnamefont {D.}~\bibnamefont {Poulin}},\
  and\ \bibinfo {author} {\bibfnamefont {W.~H.}\ \bibnamefont {Zurek}},\
  }\bibfield  {title} {\bibinfo {title} {Objective properties from subjective
  quantum states: Environment as a witness},\ }\href
  {https://doi.org/10.1103/PhysRevLett.93.220401} {\bibfield  {journal}
  {\bibinfo  {journal} {Phys. Rev. Lett.}\ }\textbf {\bibinfo {volume} {93}},\
  \bibinfo {pages} {220401} (\bibinfo {year} {2004})}\BibitemShut {NoStop}%
\bibitem [{\citenamefont {Zurek}(2009)}]{zurek-QD}%
  \BibitemOpen
  \bibfield  {author} {\bibinfo {author} {\bibfnamefont {W.~H.}\ \bibnamefont
  {Zurek}},\ }\bibfield  {title} {\bibinfo {title} {Quantum darwinism},\ }\href
  {https://doi.org/10.1038/nphys1202} {\bibfield  {journal} {\bibinfo
  {journal} {Nature Physics}\ }\textbf {\bibinfo {volume} {5}},\ \bibinfo
  {pages} {181} (\bibinfo {year} {2009})}\BibitemShut {NoStop}%
\bibitem [{\citenamefont {Ciampini}\ \emph {et~al.}(2018)\citenamefont
  {Ciampini}, \citenamefont {Pinna}, \citenamefont {Mataloni},\ and\
  \citenamefont {Paternostro}}]{paternostro18-darwin}%
  \BibitemOpen
  \bibfield  {author} {\bibinfo {author} {\bibfnamefont {M.~A.}\ \bibnamefont
  {Ciampini}}, \bibinfo {author} {\bibfnamefont {G.}~\bibnamefont {Pinna}},
  \bibinfo {author} {\bibfnamefont {P.}~\bibnamefont {Mataloni}},\ and\
  \bibinfo {author} {\bibfnamefont {M.}~\bibnamefont {Paternostro}},\
  }\bibfield  {title} {\bibinfo {title} {Experimental signature of quantum
  darwinism in photonic cluster states},\ }\href
  {https://doi.org/10.1103/PhysRevA.98.020101} {\bibfield  {journal} {\bibinfo
  {journal} {Phys. Rev. A}\ }\textbf {\bibinfo {volume} {98}},\ \bibinfo
  {pages} {020101} (\bibinfo {year} {2018})}\BibitemShut {NoStop}%
\bibitem [{\citenamefont {Unden}\ \emph {et~al.}(2019)\citenamefont {Unden},
  \citenamefont {Louzon}, \citenamefont {Zwolak}, \citenamefont {Zurek},\ and\
  \citenamefont {Jelezko}}]{unden19-darwin-exp}%
  \BibitemOpen
  \bibfield  {author} {\bibinfo {author} {\bibfnamefont {T.~K.}\ \bibnamefont
  {Unden}}, \bibinfo {author} {\bibfnamefont {D.}~\bibnamefont {Louzon}},
  \bibinfo {author} {\bibfnamefont {M.}~\bibnamefont {Zwolak}}, \bibinfo
  {author} {\bibfnamefont {W.~H.}\ \bibnamefont {Zurek}},\ and\ \bibinfo
  {author} {\bibfnamefont {F.}~\bibnamefont {Jelezko}},\ }\bibfield  {title}
  {\bibinfo {title} {Revealing the emergence of classicality using
  nitrogen-vacancy centers},\ }\href
  {https://doi.org/10.1103/PhysRevLett.123.140402} {\bibfield  {journal}
  {\bibinfo  {journal} {Phys. Rev. Lett.}\ }\textbf {\bibinfo {volume} {123}},\
  \bibinfo {pages} {140402} (\bibinfo {year} {2019})}\BibitemShut {NoStop}%
\bibitem [{\citenamefont {Zurek}(2022)}]{zurek-review}%
  \BibitemOpen
  \bibfield  {author} {\bibinfo {author} {\bibfnamefont {W.~H.}\ \bibnamefont
  {Zurek}},\ }\bibfield  {title} {\bibinfo {title} {Quantum theory of the
  classical: Einselection, envariance, quantum darwinism and extantons},\
  }\bibfield  {journal} {\bibinfo  {journal} {Entropy}\ }\textbf {\bibinfo
  {volume} {24}},\ \href {https://doi.org/10.3390/e24111520}
  {10.3390/e24111520} (\bibinfo {year} {2022})\BibitemShut {NoStop}%
\bibitem [{\citenamefont {Korbicz}\ \emph {et~al.}(2014)\citenamefont
  {Korbicz}, \citenamefont {Horodecki},\ and\ \citenamefont
  {Horodecki}}]{Korbicz-prl}%
  \BibitemOpen
  \bibfield  {author} {\bibinfo {author} {\bibfnamefont {J.~K.}\ \bibnamefont
  {Korbicz}}, \bibinfo {author} {\bibfnamefont {P.}~\bibnamefont {Horodecki}},\
  and\ \bibinfo {author} {\bibfnamefont {R.}~\bibnamefont {Horodecki}},\
  }\bibfield  {title} {\bibinfo {title} {Objectivity in a noisy photonic
  environment through quantum state information broadcasting},\ }\href
  {https://doi.org/10.1103/PhysRevLett.112.120402} {\bibfield  {journal}
  {\bibinfo  {journal} {Phys. Rev. Lett.}\ }\textbf {\bibinfo {volume} {112}},\
  \bibinfo {pages} {120402} (\bibinfo {year} {2014})}\BibitemShut {NoStop}%
\bibitem [{\citenamefont {Le}\ and\ \citenamefont
  {Olaya-Castro}(2018)}]{le-olaya-pra}%
  \BibitemOpen
  \bibfield  {author} {\bibinfo {author} {\bibfnamefont {T.~P.}\ \bibnamefont
  {Le}}\ and\ \bibinfo {author} {\bibfnamefont {A.}~\bibnamefont
  {Olaya-Castro}},\ }\bibfield  {title} {\bibinfo {title} {Objectivity (or lack
  thereof): Comparison between predictions of quantum darwinism and spectrum
  broadcast structure},\ }\href {https://doi.org/10.1103/PhysRevA.98.032103}
  {\bibfield  {journal} {\bibinfo  {journal} {Phys. Rev. A}\ }\textbf {\bibinfo
  {volume} {98}},\ \bibinfo {pages} {032103} (\bibinfo {year}
  {2018})}\BibitemShut {NoStop}%
\bibitem [{\citenamefont {Le}\ and\ \citenamefont
  {Olaya-Castro}(2019)}]{le-prl19}%
  \BibitemOpen
  \bibfield  {author} {\bibinfo {author} {\bibfnamefont {T.~P.}\ \bibnamefont
  {Le}}\ and\ \bibinfo {author} {\bibfnamefont {A.}~\bibnamefont
  {Olaya-Castro}},\ }\bibfield  {title} {\bibinfo {title} {Strong quantum
  darwinism and strong independence are equivalent to spectrum broadcast
  structure},\ }\href {https://doi.org/10.1103/PhysRevLett.122.010403}
  {\bibfield  {journal} {\bibinfo  {journal} {Phys. Rev. Lett.}\ }\textbf
  {\bibinfo {volume} {122}},\ \bibinfo {pages} {010403} (\bibinfo {year}
  {2019})}\BibitemShut {NoStop}%
\bibitem [{\citenamefont {Korbicz}(2021)}]{Korbicz-rev}%
  \BibitemOpen
  \bibfield  {author} {\bibinfo {author} {\bibfnamefont {J.~K.}\ \bibnamefont
  {Korbicz}},\ }\bibfield  {title} {\bibinfo {title} {Roads to objectivity:
  {Q}uantum {D}arwinism, {S}pectrum {B}roadcast {S}tructures, and {S}trong
  quantum {D}arwinism – a review},\ }\href
  {https://doi.org/10.22331/q-2021-11-08-571} {\bibfield  {journal} {\bibinfo
  {journal} {{Quantum}}\ }\textbf {\bibinfo {volume} {5}},\ \bibinfo {pages}
  {571} (\bibinfo {year} {2021})}\BibitemShut {NoStop}%
\bibitem [{\citenamefont {Riedel}\ \emph {et~al.}(2012)\citenamefont {Riedel},
  \citenamefont {Zurek},\ and\ \citenamefont {Zwolak}}]{riedel2012}%
  \BibitemOpen
  \bibfield  {author} {\bibinfo {author} {\bibfnamefont {C.~J.}\ \bibnamefont
  {Riedel}}, \bibinfo {author} {\bibfnamefont {W.~H.}\ \bibnamefont {Zurek}},\
  and\ \bibinfo {author} {\bibfnamefont {M.}~\bibnamefont {Zwolak}},\
  }\bibfield  {title} {\bibinfo {title} {The rise and fall of redundancy in
  decoherence and quantum darwinism},\ }\href
  {https://doi.org/10.1088/1367-2630/14/8/083010} {\bibfield  {journal}
  {\bibinfo  {journal} {New Journal of Physics}\ }\textbf {\bibinfo {volume}
  {14}},\ \bibinfo {pages} {083010} (\bibinfo {year} {2012})}\BibitemShut
  {NoStop}%
\bibitem [{\citenamefont {Campbell}\ \emph {et~al.}(2019)\citenamefont
  {Campbell}, \citenamefont {{\c{C}}akmak}, \citenamefont
  {M{\"u}stecapl{\i}o{\u{g}}lu}, \citenamefont {Paternostro},\ and\
  \citenamefont {Vacchini}}]{campbell}%
  \BibitemOpen
  \bibfield  {author} {\bibinfo {author} {\bibfnamefont {S.}~\bibnamefont
  {Campbell}}, \bibinfo {author} {\bibfnamefont {B.}~\bibnamefont
  {{\c{C}}akmak}}, \bibinfo {author} {\bibfnamefont {{\"O}.~E.}\ \bibnamefont
  {M{\"u}stecapl{\i}o{\u{g}}lu}}, \bibinfo {author} {\bibfnamefont
  {M.}~\bibnamefont {Paternostro}},\ and\ \bibinfo {author} {\bibfnamefont
  {B.}~\bibnamefont {Vacchini}},\ }\bibfield  {title} {\bibinfo {title}
  {Collisional unfolding of quantum darwinism},\ }\href
  {https://doi.org/10.1103/PhysRevA.99.042103} {\bibfield  {journal} {\bibinfo
  {journal} {Phys. Rev. A}\ }\textbf {\bibinfo {volume} {99}},\ \bibinfo
  {pages} {042103} (\bibinfo {year} {2019})}\BibitemShut {NoStop}%
\bibitem [{\citenamefont {Skinner}\ \emph {et~al.}(2019)\citenamefont
  {Skinner}, \citenamefont {Ruhman},\ and\ \citenamefont {Nahum}}]{skinner19}%
  \BibitemOpen
  \bibfield  {author} {\bibinfo {author} {\bibfnamefont {B.}~\bibnamefont
  {Skinner}}, \bibinfo {author} {\bibfnamefont {J.}~\bibnamefont {Ruhman}},\
  and\ \bibinfo {author} {\bibfnamefont {A.}~\bibnamefont {Nahum}},\ }\bibfield
   {title} {\bibinfo {title} {Measurement-induced phase transitions in the
  dynamics of entanglement},\ }\href
  {https://doi.org/10.1103/PhysRevX.9.031009} {\bibfield  {journal} {\bibinfo
  {journal} {Phys. Rev. X}\ }\textbf {\bibinfo {volume} {9}},\ \bibinfo {pages}
  {031009} (\bibinfo {year} {2019})}\BibitemShut {NoStop}%
\bibitem [{\citenamefont {Li}\ \emph {et~al.}(2018)\citenamefont {Li},
  \citenamefont {Chen},\ and\ \citenamefont {Fisher}}]{li-fisher}%
  \BibitemOpen
  \bibfield  {author} {\bibinfo {author} {\bibfnamefont {Y.}~\bibnamefont
  {Li}}, \bibinfo {author} {\bibfnamefont {X.}~\bibnamefont {Chen}},\ and\
  \bibinfo {author} {\bibfnamefont {M.~P.~A.}\ \bibnamefont {Fisher}},\
  }\bibfield  {title} {\bibinfo {title} {Quantum zeno effect and the many-body
  entanglement transition},\ }\href
  {https://doi.org/10.1103/PhysRevB.98.205136} {\bibfield  {journal} {\bibinfo
  {journal} {Phys. Rev. B}\ }\textbf {\bibinfo {volume} {98}},\ \bibinfo
  {pages} {205136} (\bibinfo {year} {2018})}\BibitemShut {NoStop}%
\bibitem [{\citenamefont {Chan}\ \emph {et~al.}(2019)\citenamefont {Chan},
  \citenamefont {Nandkishore}, \citenamefont {Pretko},\ and\ \citenamefont
  {Smith}}]{amoschan}%
  \BibitemOpen
  \bibfield  {author} {\bibinfo {author} {\bibfnamefont {A.}~\bibnamefont
  {Chan}}, \bibinfo {author} {\bibfnamefont {R.~M.}\ \bibnamefont
  {Nandkishore}}, \bibinfo {author} {\bibfnamefont {M.}~\bibnamefont
  {Pretko}},\ and\ \bibinfo {author} {\bibfnamefont {G.}~\bibnamefont
  {Smith}},\ }\bibfield  {title} {\bibinfo {title} {Unitary-projective
  entanglement dynamics},\ }\href {https://doi.org/10.1103/PhysRevB.99.224307}
  {\bibfield  {journal} {\bibinfo  {journal} {Phys. Rev. B}\ }\textbf {\bibinfo
  {volume} {99}},\ \bibinfo {pages} {224307} (\bibinfo {year}
  {2019})}\BibitemShut {NoStop}%
\bibitem [{\citenamefont {Jian}\ \emph {et~al.}(2020)\citenamefont {Jian},
  \citenamefont {You}, \citenamefont {Vasseur},\ and\ \citenamefont
  {Ludwig}}]{vasseur-ludwig}%
  \BibitemOpen
  \bibfield  {author} {\bibinfo {author} {\bibfnamefont {C.-M.}\ \bibnamefont
  {Jian}}, \bibinfo {author} {\bibfnamefont {Y.-Z.}\ \bibnamefont {You}},
  \bibinfo {author} {\bibfnamefont {R.}~\bibnamefont {Vasseur}},\ and\ \bibinfo
  {author} {\bibfnamefont {A.~W.~W.}\ \bibnamefont {Ludwig}},\ }\bibfield
  {title} {\bibinfo {title} {Measurement-induced criticality in random quantum
  circuits},\ }\href {https://doi.org/10.1103/PhysRevB.101.104302} {\bibfield
  {journal} {\bibinfo  {journal} {Phys. Rev. B}\ }\textbf {\bibinfo {volume}
  {101}},\ \bibinfo {pages} {104302} (\bibinfo {year} {2020})}\BibitemShut
  {NoStop}%
\bibitem [{\citenamefont {Cao}\ \emph {et~al.}(2019)\citenamefont {Cao},
  \citenamefont {Tilloy},\ and\ \citenamefont {De~Luca}}]{cao-deluca}%
  \BibitemOpen
  \bibfield  {author} {\bibinfo {author} {\bibfnamefont {X.}~\bibnamefont
  {Cao}}, \bibinfo {author} {\bibfnamefont {A.}~\bibnamefont {Tilloy}},\ and\
  \bibinfo {author} {\bibfnamefont {A.}~\bibnamefont {De~Luca}},\ }\bibfield
  {title} {\bibinfo {title} {{Entanglement in a fermion chain under continuous
  monitoring}},\ }\href {https://doi.org/10.21468/SciPostPhys.7.2.024}
  {\bibfield  {journal} {\bibinfo  {journal} {SciPost Phys.}\ }\textbf
  {\bibinfo {volume} {7}},\ \bibinfo {pages} {024} (\bibinfo {year}
  {2019})}\BibitemShut {NoStop}%
\bibitem [{\citenamefont {Szyniszewski}\ \emph {et~al.}(2019)\citenamefont
  {Szyniszewski}, \citenamefont {Romito},\ and\ \citenamefont
  {Schomerus}}]{schomerus19}%
  \BibitemOpen
  \bibfield  {author} {\bibinfo {author} {\bibfnamefont {M.}~\bibnamefont
  {Szyniszewski}}, \bibinfo {author} {\bibfnamefont {A.}~\bibnamefont
  {Romito}},\ and\ \bibinfo {author} {\bibfnamefont {H.}~\bibnamefont
  {Schomerus}},\ }\bibfield  {title} {\bibinfo {title} {Entanglement transition
  from variable-strength weak measurements},\ }\href
  {https://doi.org/10.1103/PhysRevB.100.064204} {\bibfield  {journal} {\bibinfo
   {journal} {Phys. Rev. B}\ }\textbf {\bibinfo {volume} {100}},\ \bibinfo
  {pages} {064204} (\bibinfo {year} {2019})}\BibitemShut {NoStop}%
\bibitem [{\citenamefont {Li}\ \emph {et~al.}(2019)\citenamefont {Li},
  \citenamefont {Chen},\ and\ \citenamefont {Fisher}}]{li-fisher2}%
  \BibitemOpen
  \bibfield  {author} {\bibinfo {author} {\bibfnamefont {Y.}~\bibnamefont
  {Li}}, \bibinfo {author} {\bibfnamefont {X.}~\bibnamefont {Chen}},\ and\
  \bibinfo {author} {\bibfnamefont {M.~P.~A.}\ \bibnamefont {Fisher}},\
  }\bibfield  {title} {\bibinfo {title} {Measurement-driven entanglement
  transition in hybrid quantum circuits},\ }\href
  {https://doi.org/10.1103/PhysRevB.100.134306} {\bibfield  {journal} {\bibinfo
   {journal} {Phys. Rev. B}\ }\textbf {\bibinfo {volume} {100}},\ \bibinfo
  {pages} {134306} (\bibinfo {year} {2019})}\BibitemShut {NoStop}%
\bibitem [{\citenamefont {Choi}\ \emph {et~al.}(2020)\citenamefont {Choi},
  \citenamefont {Bao}, \citenamefont {Qi},\ and\ \citenamefont
  {Altman}}]{choi-altman-prl}%
  \BibitemOpen
  \bibfield  {author} {\bibinfo {author} {\bibfnamefont {S.}~\bibnamefont
  {Choi}}, \bibinfo {author} {\bibfnamefont {Y.}~\bibnamefont {Bao}}, \bibinfo
  {author} {\bibfnamefont {X.-L.}\ \bibnamefont {Qi}},\ and\ \bibinfo {author}
  {\bibfnamefont {E.}~\bibnamefont {Altman}},\ }\bibfield  {title} {\bibinfo
  {title} {Quantum error correction in scrambling dynamics and
  measurement-induced phase transition},\ }\href
  {https://doi.org/10.1103/PhysRevLett.125.030505} {\bibfield  {journal}
  {\bibinfo  {journal} {Phys. Rev. Lett.}\ }\textbf {\bibinfo {volume} {125}},\
  \bibinfo {pages} {030505} (\bibinfo {year} {2020})}\BibitemShut {NoStop}%
\bibitem [{\citenamefont {Bao}\ \emph {et~al.}(2020)\citenamefont {Bao},
  \citenamefont {Choi},\ and\ \citenamefont {Altman}}]{bao-altman}%
  \BibitemOpen
  \bibfield  {author} {\bibinfo {author} {\bibfnamefont {Y.}~\bibnamefont
  {Bao}}, \bibinfo {author} {\bibfnamefont {S.}~\bibnamefont {Choi}},\ and\
  \bibinfo {author} {\bibfnamefont {E.}~\bibnamefont {Altman}},\ }\bibfield
  {title} {\bibinfo {title} {Theory of the phase transition in random unitary
  circuits with measurements},\ }\href
  {https://doi.org/10.1103/PhysRevB.101.104301} {\bibfield  {journal} {\bibinfo
   {journal} {Phys. Rev. B}\ }\textbf {\bibinfo {volume} {101}},\ \bibinfo
  {pages} {104301} (\bibinfo {year} {2020})}\BibitemShut {NoStop}%
\bibitem [{\citenamefont {Gullans}\ and\ \citenamefont
  {Huse}(2020{\natexlab{a}})}]{gullans-huse-prx}%
  \BibitemOpen
  \bibfield  {author} {\bibinfo {author} {\bibfnamefont {M.~J.}\ \bibnamefont
  {Gullans}}\ and\ \bibinfo {author} {\bibfnamefont {D.~A.}\ \bibnamefont
  {Huse}},\ }\bibfield  {title} {\bibinfo {title} {Dynamical purification phase
  transition induced by quantum measurements},\ }\href
  {https://doi.org/10.1103/PhysRevX.10.041020} {\bibfield  {journal} {\bibinfo
  {journal} {Phys. Rev. X}\ }\textbf {\bibinfo {volume} {10}},\ \bibinfo
  {pages} {041020} (\bibinfo {year} {2020}{\natexlab{a}})}\BibitemShut
  {NoStop}%
\bibitem [{\citenamefont {Turkeshi}\ \emph {et~al.}(2021)\citenamefont
  {Turkeshi}, \citenamefont {Biella}, \citenamefont {Fazio}, \citenamefont
  {Dalmonte},\ and\ \citenamefont {Schir\'o}}]{schiro}%
  \BibitemOpen
  \bibfield  {author} {\bibinfo {author} {\bibfnamefont {X.}~\bibnamefont
  {Turkeshi}}, \bibinfo {author} {\bibfnamefont {A.}~\bibnamefont {Biella}},
  \bibinfo {author} {\bibfnamefont {R.}~\bibnamefont {Fazio}}, \bibinfo
  {author} {\bibfnamefont {M.}~\bibnamefont {Dalmonte}},\ and\ \bibinfo
  {author} {\bibfnamefont {M.}~\bibnamefont {Schir\'o}},\ }\bibfield  {title}
  {\bibinfo {title} {Measurement-induced entanglement transitions in the
  quantum ising chain: From infinite to zero clicks},\ }\href
  {https://doi.org/10.1103/PhysRevB.103.224210} {\bibfield  {journal} {\bibinfo
   {journal} {Phys. Rev. B}\ }\textbf {\bibinfo {volume} {103}},\ \bibinfo
  {pages} {224210} (\bibinfo {year} {2021})}\BibitemShut {NoStop}%
\bibitem [{\citenamefont {Fisher}\ \emph {et~al.}(2023)\citenamefont {Fisher},
  \citenamefont {Khemani}, \citenamefont {Nahum},\ and\ \citenamefont
  {Vijay}}]{MIPTrev}%
  \BibitemOpen
  \bibfield  {author} {\bibinfo {author} {\bibfnamefont {M.~P.}\ \bibnamefont
  {Fisher}}, \bibinfo {author} {\bibfnamefont {V.}~\bibnamefont {Khemani}},
  \bibinfo {author} {\bibfnamefont {A.}~\bibnamefont {Nahum}},\ and\ \bibinfo
  {author} {\bibfnamefont {S.}~\bibnamefont {Vijay}},\ }\bibfield  {title}
  {\bibinfo {title} {Random quantum circuits},\ }\href
  {https://doi.org/10.1146/annurev-conmatphys-031720-030658} {\bibfield
  {journal} {\bibinfo  {journal} {Annual Review of Condensed Matter Physics}\
  }\textbf {\bibinfo {volume} {14}},\ \bibinfo {pages} {335} (\bibinfo {year}
  {2023})}\BibitemShut {NoStop}%
\bibitem [{\citenamefont {Li}\ \emph {et~al.}(2022)\citenamefont {Li},
  \citenamefont {Zou}, \citenamefont {Glorioso}, \citenamefont {Altman},\ and\
  \citenamefont {Fisher}}]{li2022cross}%
  \BibitemOpen
  \bibfield  {author} {\bibinfo {author} {\bibfnamefont {Y.}~\bibnamefont
  {Li}}, \bibinfo {author} {\bibfnamefont {Y.}~\bibnamefont {Zou}}, \bibinfo
  {author} {\bibfnamefont {P.}~\bibnamefont {Glorioso}}, \bibinfo {author}
  {\bibfnamefont {E.}~\bibnamefont {Altman}},\ and\ \bibinfo {author}
  {\bibfnamefont {M.}~\bibnamefont {Fisher}},\ }\bibfield  {title} {\bibinfo
  {title} {Cross entropy benchmark for measurement-induced phase transitions},\
  }\bibfield  {journal} {\bibinfo  {journal} {arXiv:2209.00609}\ }\href
  {https://doi.org/10.48550/arXiv.2209.00609} {10.48550/arXiv.2209.00609}
  (\bibinfo {year} {2022})\BibitemShut {NoStop}%
\bibitem [{\citenamefont {Ippoliti}\ and\ \citenamefont
  {Khemani}(2021)}]{khemani-post}%
  \BibitemOpen
  \bibfield  {author} {\bibinfo {author} {\bibfnamefont {M.}~\bibnamefont
  {Ippoliti}}\ and\ \bibinfo {author} {\bibfnamefont {V.}~\bibnamefont
  {Khemani}},\ }\bibfield  {title} {\bibinfo {title} {Postselection-free
  entanglement dynamics via spacetime duality},\ }\href
  {https://doi.org/10.1103/PhysRevLett.126.060501} {\bibfield  {journal}
  {\bibinfo  {journal} {Phys. Rev. Lett.}\ }\textbf {\bibinfo {volume} {126}},\
  \bibinfo {pages} {060501} (\bibinfo {year} {2021})}\BibitemShut {NoStop}%
\bibitem [{\citenamefont {Noel}\ \emph {et~al.}(2022)\citenamefont {Noel},
  \citenamefont {Niroula}, \citenamefont {Zhu}, \citenamefont {Risinger},
  \citenamefont {Egan}, \citenamefont {Biswas}, \citenamefont {Cetina},
  \citenamefont {Gorshkov}, \citenamefont {Gullans}, \citenamefont {Huse},\
  and\ \citenamefont {Monroe}}]{MIPTnature}%
  \BibitemOpen
  \bibfield  {author} {\bibinfo {author} {\bibfnamefont {C.}~\bibnamefont
  {Noel}}, \bibinfo {author} {\bibfnamefont {P.}~\bibnamefont {Niroula}},
  \bibinfo {author} {\bibfnamefont {D.}~\bibnamefont {Zhu}}, \bibinfo {author}
  {\bibfnamefont {A.}~\bibnamefont {Risinger}}, \bibinfo {author}
  {\bibfnamefont {L.}~\bibnamefont {Egan}}, \bibinfo {author} {\bibfnamefont
  {D.}~\bibnamefont {Biswas}}, \bibinfo {author} {\bibfnamefont
  {M.}~\bibnamefont {Cetina}}, \bibinfo {author} {\bibfnamefont {A.~V.}\
  \bibnamefont {Gorshkov}}, \bibinfo {author} {\bibfnamefont {M.~J.}\
  \bibnamefont {Gullans}}, \bibinfo {author} {\bibfnamefont {D.~A.}\
  \bibnamefont {Huse}},\ and\ \bibinfo {author} {\bibfnamefont
  {C.}~\bibnamefont {Monroe}},\ }\bibfield  {title} {\bibinfo {title}
  {Measurement-induced quantum phases realized in a trapped-ion quantum
  computer},\ }\href {https://doi.org/10.1038/s41567-022-01619-7} {\bibfield
  {journal} {\bibinfo  {journal} {Nature Physics}\ }\textbf {\bibinfo {volume}
  {18}},\ \bibinfo {pages} {760} (\bibinfo {year} {2022})}\BibitemShut
  {NoStop}%
\bibitem [{\citenamefont {Touil}\ \emph {et~al.}(2022)\citenamefont {Touil},
  \citenamefont {Yan}, \citenamefont {Girolami}, \citenamefont {Deffner},\ and\
  \citenamefont {Zurek}}]{girolami}%
  \BibitemOpen
  \bibfield  {author} {\bibinfo {author} {\bibfnamefont {A.}~\bibnamefont
  {Touil}}, \bibinfo {author} {\bibfnamefont {B.}~\bibnamefont {Yan}}, \bibinfo
  {author} {\bibfnamefont {D.}~\bibnamefont {Girolami}}, \bibinfo {author}
  {\bibfnamefont {S.}~\bibnamefont {Deffner}},\ and\ \bibinfo {author}
  {\bibfnamefont {W.~H.}\ \bibnamefont {Zurek}},\ }\bibfield  {title} {\bibinfo
  {title} {Eavesdropping on the decohering environment: Quantum darwinism,
  amplification, and the origin of objective classical reality},\ }\href
  {https://doi.org/10.1103/PhysRevLett.128.010401} {\bibfield  {journal}
  {\bibinfo  {journal} {Phys. Rev. Lett.}\ }\textbf {\bibinfo {volume} {128}},\
  \bibinfo {pages} {010401} (\bibinfo {year} {2022})}\BibitemShut {NoStop}%
\bibitem [{\citenamefont {Gullans}\ and\ \citenamefont
  {Huse}(2020{\natexlab{b}})}]{gullans-huse-prl}%
  \BibitemOpen
  \bibfield  {author} {\bibinfo {author} {\bibfnamefont {M.~J.}\ \bibnamefont
  {Gullans}}\ and\ \bibinfo {author} {\bibfnamefont {D.~A.}\ \bibnamefont
  {Huse}},\ }\bibfield  {title} {\bibinfo {title} {Scalable probes of
  measurement-induced criticality},\ }\href
  {https://doi.org/10.1103/PhysRevLett.125.070606} {\bibfield  {journal}
  {\bibinfo  {journal} {Phys. Rev. Lett.}\ }\textbf {\bibinfo {volume} {125}},\
  \bibinfo {pages} {070606} (\bibinfo {year} {2020}{\natexlab{b}})}\BibitemShut
  {NoStop}%
\bibitem [{\citenamefont {Fan}\ \emph {et~al.}(2021)\citenamefont {Fan},
  \citenamefont {Vijay}, \citenamefont {Vishwanath},\ and\ \citenamefont
  {You}}]{Vijay-Vishwanath}%
  \BibitemOpen
  \bibfield  {author} {\bibinfo {author} {\bibfnamefont {R.}~\bibnamefont
  {Fan}}, \bibinfo {author} {\bibfnamefont {S.}~\bibnamefont {Vijay}}, \bibinfo
  {author} {\bibfnamefont {A.}~\bibnamefont {Vishwanath}},\ and\ \bibinfo
  {author} {\bibfnamefont {Y.-Z.}\ \bibnamefont {You}},\ }\bibfield  {title}
  {\bibinfo {title} {Self-organized error correction in random unitary circuits
  with measurement},\ }\href {https://doi.org/10.1103/PhysRevB.103.174309}
  {\bibfield  {journal} {\bibinfo  {journal} {Phys. Rev. B}\ }\textbf {\bibinfo
  {volume} {103}},\ \bibinfo {pages} {174309} (\bibinfo {year}
  {2021})}\BibitemShut {NoStop}%
\bibitem [{\citenamefont {Li}\ and\ \citenamefont
  {Fisher}(2021)}]{li-fisher-QEC}%
  \BibitemOpen
  \bibfield  {author} {\bibinfo {author} {\bibfnamefont {Y.}~\bibnamefont
  {Li}}\ and\ \bibinfo {author} {\bibfnamefont {M.~P.~A.}\ \bibnamefont
  {Fisher}},\ }\bibfield  {title} {\bibinfo {title} {Statistical mechanics of
  quantum error correcting codes},\ }\href
  {https://doi.org/10.1103/PhysRevB.103.104306} {\bibfield  {journal} {\bibinfo
   {journal} {Phys. Rev. B}\ }\textbf {\bibinfo {volume} {103}},\ \bibinfo
  {pages} {104306} (\bibinfo {year} {2021})}\BibitemShut {NoStop}%
\bibitem [{\citenamefont {Lovas}\ \emph {et~al.}(2023)\citenamefont {Lovas},
  \citenamefont {Agrawal},\ and\ \citenamefont {Vijay}}]{vijay-encoding}%
  \BibitemOpen
  \bibfield  {author} {\bibinfo {author} {\bibfnamefont {I.}~\bibnamefont
  {Lovas}}, \bibinfo {author} {\bibfnamefont {U.}~\bibnamefont {Agrawal}},\
  and\ \bibinfo {author} {\bibfnamefont {S.}~\bibnamefont {Vijay}},\ }\bibfield
   {title} {\bibinfo {title} {Quantum coding transitions in the presence of
  boundary dissipation},\ }\bibfield  {journal} {\bibinfo  {journal}
  {Preprint}\ }\href {https://doi.org/10.48550/arXiv.2304.02664}
  {10.48550/arXiv.2304.02664} (\bibinfo {year} {2023})\BibitemShut {NoStop}%
\bibitem [{\citenamefont {Nahum}\ \emph {et~al.}(2021)\citenamefont {Nahum},
  \citenamefont {Roy}, \citenamefont {Skinner},\ and\ \citenamefont
  {Ruhman}}]{nahum21}%
  \BibitemOpen
  \bibfield  {author} {\bibinfo {author} {\bibfnamefont {A.}~\bibnamefont
  {Nahum}}, \bibinfo {author} {\bibfnamefont {S.}~\bibnamefont {Roy}}, \bibinfo
  {author} {\bibfnamefont {B.}~\bibnamefont {Skinner}},\ and\ \bibinfo {author}
  {\bibfnamefont {J.}~\bibnamefont {Ruhman}},\ }\bibfield  {title} {\bibinfo
  {title} {Measurement and entanglement phase transitions in all-to-all quantum
  circuits, on quantum trees, and in landau-ginsburg theory},\ }\href
  {https://doi.org/10.1103/PRXQuantum.2.010352} {\bibfield  {journal} {\bibinfo
   {journal} {PRX Quantum}\ }\textbf {\bibinfo {volume} {2}},\ \bibinfo {pages}
  {010352} (\bibinfo {year} {2021})}\BibitemShut {NoStop}%
\bibitem [{\citenamefont {Feng}\ \emph {et~al.}(2023)\citenamefont {Feng},
  \citenamefont {Skinner},\ and\ \citenamefont {Nahum}}]{feng2022measurement}%
  \BibitemOpen
  \bibfield  {author} {\bibinfo {author} {\bibfnamefont {X.}~\bibnamefont
  {Feng}}, \bibinfo {author} {\bibfnamefont {B.}~\bibnamefont {Skinner}},\ and\
  \bibinfo {author} {\bibfnamefont {A.}~\bibnamefont {Nahum}},\ }\bibfield
  {title} {\bibinfo {title} {Measurement-induced phase transitions on dynamical
  quantum trees},\ }\href {https://doi.org/10.1103/PRXQuantum.4.030333}
  {\bibfield  {journal} {\bibinfo  {journal} {PRX Quantum}\ }\textbf {\bibinfo
  {volume} {4}},\ \bibinfo {pages} {030333} (\bibinfo {year}
  {2023})}\BibitemShut {NoStop}%
\bibitem [{\citenamefont {Gottesman}(1998)}]{gottesman1998heisenberg}%
  \BibitemOpen
  \bibfield  {author} {\bibinfo {author} {\bibfnamefont {D.}~\bibnamefont
  {Gottesman}},\ }\bibfield  {title} {\bibinfo {title} {The heisenberg
  representation of quantum computers},\ }\bibfield  {journal} {\bibinfo
  {journal} {Preprint}\ }\href
  {https://doi.org/10.48550/arXiv.quant-ph/9807006}
  {10.48550/arXiv.quant-ph/9807006} (\bibinfo {year} {1998})\BibitemShut
  {NoStop}%
\bibitem [{\citenamefont {Aaronson}\ and\ \citenamefont
  {Gottesman}(2004)}]{aaronson-gottesman}%
  \BibitemOpen
  \bibfield  {author} {\bibinfo {author} {\bibfnamefont {S.}~\bibnamefont
  {Aaronson}}\ and\ \bibinfo {author} {\bibfnamefont {D.}~\bibnamefont
  {Gottesman}},\ }\bibfield  {title} {\bibinfo {title} {Improved simulation of
  stabilizer circuits},\ }\href {https://doi.org/10.1103/PhysRevA.70.052328}
  {\bibfield  {journal} {\bibinfo  {journal} {Phys. Rev. A}\ }\textbf {\bibinfo
  {volume} {70}},\ \bibinfo {pages} {052328} (\bibinfo {year}
  {2004})}\BibitemShut {NoStop}%
\bibitem [{sup()}]{supp}%
  \BibitemOpen
  \href@noop {} {\bibinfo {title} {{See Supplemental Material.}}}\BibitemShut
  {Stop}%
\bibitem [{\citenamefont {Zhou}\ and\ \citenamefont {Nahum}(2019)}]{tianci}%
  \BibitemOpen
  \bibfield  {author} {\bibinfo {author} {\bibfnamefont {T.}~\bibnamefont
  {Zhou}}\ and\ \bibinfo {author} {\bibfnamefont {A.}~\bibnamefont {Nahum}},\
  }\bibfield  {title} {\bibinfo {title} {Emergent statistical mechanics of
  entanglement in random unitary circuits},\ }\href
  {https://doi.org/10.1103/PhysRevB.99.174205} {\bibfield  {journal} {\bibinfo
  {journal} {Phys. Rev. B}\ }\textbf {\bibinfo {volume} {99}},\ \bibinfo
  {pages} {174205} (\bibinfo {year} {2019})}\BibitemShut {NoStop}%
\bibitem [{inp()}]{inprep}%
  \BibitemOpen
  \href@noop {} {\bibinfo {title} {{Beno\^it Fert\'e, Xiangyu Cao, in
  preparation.}}}\BibitemShut {Stop}%
\bibitem [{\citenamefont {Page}(1993)}]{page}%
  \BibitemOpen
  \bibfield  {author} {\bibinfo {author} {\bibfnamefont {D.~N.}\ \bibnamefont
  {Page}},\ }\bibfield  {title} {\bibinfo {title} {Average entropy of a
  subsystem},\ }\href {https://doi.org/10.1103/PhysRevLett.71.1291} {\bibfield
  {journal} {\bibinfo  {journal} {Phys. Rev. Lett.}\ }\textbf {\bibinfo
  {volume} {71}},\ \bibinfo {pages} {1291} (\bibinfo {year}
  {1993})}\BibitemShut {NoStop}%
\bibitem [{\citenamefont {Ollivier}\ and\ \citenamefont
  {Zurek}(2001)}]{discord-zurek}%
  \BibitemOpen
  \bibfield  {author} {\bibinfo {author} {\bibfnamefont {H.}~\bibnamefont
  {Ollivier}}\ and\ \bibinfo {author} {\bibfnamefont {W.~H.}\ \bibnamefont
  {Zurek}},\ }\bibfield  {title} {\bibinfo {title} {Quantum discord: A measure
  of the quantumness of correlations},\ }\href
  {https://doi.org/10.1103/PhysRevLett.88.017901} {\bibfield  {journal}
  {\bibinfo  {journal} {Phys. Rev. Lett.}\ }\textbf {\bibinfo {volume} {88}},\
  \bibinfo {pages} {017901} (\bibinfo {year} {2001})}\BibitemShut {NoStop}%
\bibitem [{\citenamefont {Henderson}\ and\ \citenamefont
  {Vedral}(2001)}]{Henderson_2001}%
  \BibitemOpen
  \bibfield  {author} {\bibinfo {author} {\bibfnamefont {L.}~\bibnamefont
  {Henderson}}\ and\ \bibinfo {author} {\bibfnamefont {V.}~\bibnamefont
  {Vedral}},\ }\bibfield  {title} {\bibinfo {title} {Classical, quantum and
  total correlations},\ }\href {https://doi.org/10.1088/0305-4470/34/35/315}
  {\bibfield  {journal} {\bibinfo  {journal} {Journal of Physics A:
  Mathematical and General}\ }\textbf {\bibinfo {volume} {34}},\ \bibinfo
  {pages} {6899} (\bibinfo {year} {2001})}\BibitemShut {NoStop}%
\bibitem [{\citenamefont {Holzhey}\ \emph {et~al.}(1994)\citenamefont
  {Holzhey}, \citenamefont {Larsen},\ and\ \citenamefont {Wilczek}}]{wilczek}%
  \BibitemOpen
  \bibfield  {author} {\bibinfo {author} {\bibfnamefont {C.}~\bibnamefont
  {Holzhey}}, \bibinfo {author} {\bibfnamefont {F.}~\bibnamefont {Larsen}},\
  and\ \bibinfo {author} {\bibfnamefont {F.}~\bibnamefont {Wilczek}},\
  }\bibfield  {title} {\bibinfo {title} {Geometric and renormalized entropy in
  conformal field theory},\ }\href
  {https://doi.org/https://doi.org/10.1016/0550-3213(94)90402-2} {\bibfield
  {journal} {\bibinfo  {journal} {Nuclear Physics B}\ }\textbf {\bibinfo
  {volume} {424}},\ \bibinfo {pages} {443} (\bibinfo {year}
  {1994})}\BibitemShut {NoStop}%
\bibitem [{\citenamefont {Calabrese}\ and\ \citenamefont
  {Cardy}(2004)}]{CalabreseCardy_2004}%
  \BibitemOpen
  \bibfield  {author} {\bibinfo {author} {\bibfnamefont {P.}~\bibnamefont
  {Calabrese}}\ and\ \bibinfo {author} {\bibfnamefont {J.}~\bibnamefont
  {Cardy}},\ }\bibfield  {title} {\bibinfo {title} {Entanglement entropy and
  quantum field theory},\ }\href
  {https://doi.org/10.1088/1742-5468/2004/06/P06002} {\bibfield  {journal}
  {\bibinfo  {journal} {Journal of Statistical Mechanics: Theory and
  Experiment}\ }\textbf {\bibinfo {volume} {2004}},\ \bibinfo {pages} {P06002}
  (\bibinfo {year} {2004})}\BibitemShut {NoStop}%
\bibitem [{\citenamefont {Calabrese}\ \emph {et~al.}(2012)\citenamefont
  {Calabrese}, \citenamefont {Cardy},\ and\ \citenamefont
  {Tonni}}]{tonni-negativity}%
  \BibitemOpen
  \bibfield  {author} {\bibinfo {author} {\bibfnamefont {P.}~\bibnamefont
  {Calabrese}}, \bibinfo {author} {\bibfnamefont {J.}~\bibnamefont {Cardy}},\
  and\ \bibinfo {author} {\bibfnamefont {E.}~\bibnamefont {Tonni}},\ }\bibfield
   {title} {\bibinfo {title} {Entanglement negativity in quantum field
  theory},\ }\href {https://doi.org/10.1103/PhysRevLett.109.130502} {\bibfield
  {journal} {\bibinfo  {journal} {Phys. Rev. Lett.}\ }\textbf {\bibinfo
  {volume} {109}},\ \bibinfo {pages} {130502} (\bibinfo {year}
  {2012})}\BibitemShut {NoStop}%
\bibitem [{\citenamefont {Lashkari}(2014)}]{lashkari}%
  \BibitemOpen
  \bibfield  {author} {\bibinfo {author} {\bibfnamefont {N.}~\bibnamefont
  {Lashkari}},\ }\bibfield  {title} {\bibinfo {title} {Relative entropies in
  conformal field theory},\ }\href
  {https://doi.org/10.1103/PhysRevLett.113.051602} {\bibfield  {journal}
  {\bibinfo  {journal} {Phys. Rev. Lett.}\ }\textbf {\bibinfo {volume} {113}},\
  \bibinfo {pages} {051602} (\bibinfo {year} {2014})}\BibitemShut {NoStop}%
\bibitem [{\citenamefont {Zhang}\ \emph {et~al.}(2019)\citenamefont {Zhang},
  \citenamefont {Ruggiero},\ and\ \citenamefont
  {Calabrese}}]{ruggiero-trace-dis}%
  \BibitemOpen
  \bibfield  {author} {\bibinfo {author} {\bibfnamefont {J.}~\bibnamefont
  {Zhang}}, \bibinfo {author} {\bibfnamefont {P.}~\bibnamefont {Ruggiero}},\
  and\ \bibinfo {author} {\bibfnamefont {P.}~\bibnamefont {Calabrese}},\
  }\bibfield  {title} {\bibinfo {title} {Subsystem trace distance in quantum
  field theory},\ }\href {https://doi.org/10.1103/PhysRevLett.122.141602}
  {\bibfield  {journal} {\bibinfo  {journal} {Phys. Rev. Lett.}\ }\textbf
  {\bibinfo {volume} {122}},\ \bibinfo {pages} {141602} (\bibinfo {year}
  {2019})}\BibitemShut {NoStop}%
\bibitem [{\citenamefont {Kudler-Flam}(2023)}]{kudler-renyiMI}%
  \BibitemOpen
  \bibfield  {author} {\bibinfo {author} {\bibfnamefont {J.}~\bibnamefont
  {Kudler-Flam}},\ }\bibfield  {title} {\bibinfo {title} {R\'enyi mutual
  information in quantum field theory},\ }\href
  {https://doi.org/10.1103/PhysRevLett.130.021603} {\bibfield  {journal}
  {\bibinfo  {journal} {Phys. Rev. Lett.}\ }\textbf {\bibinfo {volume} {130}},\
  \bibinfo {pages} {021603} (\bibinfo {year} {2023})}\BibitemShut {NoStop}%
\bibitem [{\citenamefont {Nahum}\ \emph {et~al.}(2017)\citenamefont {Nahum},
  \citenamefont {Ruhman}, \citenamefont {Vijay},\ and\ \citenamefont
  {Haah}}]{nahum-entanglement}%
  \BibitemOpen
  \bibfield  {author} {\bibinfo {author} {\bibfnamefont {A.}~\bibnamefont
  {Nahum}}, \bibinfo {author} {\bibfnamefont {J.}~\bibnamefont {Ruhman}},
  \bibinfo {author} {\bibfnamefont {S.}~\bibnamefont {Vijay}},\ and\ \bibinfo
  {author} {\bibfnamefont {J.}~\bibnamefont {Haah}},\ }\bibfield  {title}
  {\bibinfo {title} {Quantum entanglement growth under random unitary
  dynamics},\ }\href {https://doi.org/10.1103/PhysRevX.7.031016} {\bibfield
  {journal} {\bibinfo  {journal} {Phys. Rev. X}\ }\textbf {\bibinfo {volume}
  {7}},\ \bibinfo {pages} {031016} (\bibinfo {year} {2017})}\BibitemShut
  {NoStop}%
\bibitem [{\citenamefont {Nahum}\ \emph {et~al.}(2018)\citenamefont {Nahum},
  \citenamefont {Vijay},\ and\ \citenamefont {Haah}}]{nahum-otoc}%
  \BibitemOpen
  \bibfield  {author} {\bibinfo {author} {\bibfnamefont {A.}~\bibnamefont
  {Nahum}}, \bibinfo {author} {\bibfnamefont {S.}~\bibnamefont {Vijay}},\ and\
  \bibinfo {author} {\bibfnamefont {J.}~\bibnamefont {Haah}},\ }\bibfield
  {title} {\bibinfo {title} {Operator spreading in random unitary circuits},\
  }\href {https://doi.org/10.1103/PhysRevX.8.021014} {\bibfield  {journal}
  {\bibinfo  {journal} {Phys. Rev. X}\ }\textbf {\bibinfo {volume} {8}},\
  \bibinfo {pages} {021014} (\bibinfo {year} {2018})}\BibitemShut {NoStop}%
\bibitem [{\citenamefont {Collins}\ \emph {et~al.}(2022)\citenamefont
  {Collins}, \citenamefont {Matsumoto},\ and\ \citenamefont {Novak}}]{colin}%
  \BibitemOpen
  \bibfield  {author} {\bibinfo {author} {\bibfnamefont {B.}~\bibnamefont
  {Collins}}, \bibinfo {author} {\bibfnamefont {S.}~\bibnamefont {Matsumoto}},\
  and\ \bibinfo {author} {\bibfnamefont {J.}~\bibnamefont {Novak}},\ }\bibfield
   {title} {\bibinfo {title} {{The Weingarten Calculus}},\ }\bibfield
  {journal} {\bibinfo  {journal} {Notice of the AMS}\ }\textbf {\bibinfo
  {volume} {69}},\ \href {https://doi.org/10.1090/noti2474} {10.1090/noti2474}
  (\bibinfo {year} {2022})\BibitemShut {NoStop}%
\bibitem [{\citenamefont {Dankert}\ \emph {et~al.}(2009)\citenamefont
  {Dankert}, \citenamefont {Cleve}, \citenamefont {Emerson},\ and\
  \citenamefont {Livine}}]{livine-2design}%
  \BibitemOpen
  \bibfield  {author} {\bibinfo {author} {\bibfnamefont {C.}~\bibnamefont
  {Dankert}}, \bibinfo {author} {\bibfnamefont {R.}~\bibnamefont {Cleve}},
  \bibinfo {author} {\bibfnamefont {J.}~\bibnamefont {Emerson}},\ and\ \bibinfo
  {author} {\bibfnamefont {E.}~\bibnamefont {Livine}},\ }\bibfield  {title}
  {\bibinfo {title} {Exact and approximate unitary 2-designs and their
  application to fidelity estimation},\ }\href
  {https://doi.org/10.1103/PhysRevA.80.012304} {\bibfield  {journal} {\bibinfo
  {journal} {Phys. Rev. A}\ }\textbf {\bibinfo {volume} {80}},\ \bibinfo
  {pages} {012304} (\bibinfo {year} {2009})}\BibitemShut {NoStop}%
\bibitem [{\citenamefont {Greenberger}\ \emph {et~al.}(1989)\citenamefont
  {Greenberger}, \citenamefont {Horne},\ and\ \citenamefont
  {Zeilinger}}]{greenberger1989}%
  \BibitemOpen
  \bibfield  {author} {\bibinfo {author} {\bibfnamefont {D.~M.}\ \bibnamefont
  {Greenberger}}, \bibinfo {author} {\bibfnamefont {M.~A.}\ \bibnamefont
  {Horne}},\ and\ \bibinfo {author} {\bibfnamefont {A.}~\bibnamefont
  {Zeilinger}},\ }\bibinfo {title} {Going beyond bell's theorem},\ in\ \href
  {https://doi.org/10.1007/978-94-017-0849-4{\_}10} {\emph {\bibinfo
  {booktitle} {Bell's Theorem, Quantum Theory and Conceptions of the
  Universe}}},\ \bibinfo {editor} {edited by\ \bibinfo {editor} {\bibfnamefont
  {M.}~\bibnamefont {Kafatos}}}\ (\bibinfo  {publisher} {Springer
  Netherlands},\ \bibinfo {address} {Dordrecht},\ \bibinfo {year} {1989})\ pp.\
  \bibinfo {pages} {69--72}\BibitemShut {NoStop}%
\bibitem [{\citenamefont {Ryan}\ \emph {et~al.}(2021)\citenamefont {Ryan},
  \citenamefont {Paternostro},\ and\ \citenamefont {Campbell}}]{Ryan-onion}%
  \BibitemOpen
  \bibfield  {author} {\bibinfo {author} {\bibfnamefont {E.}~\bibnamefont
  {Ryan}}, \bibinfo {author} {\bibfnamefont {M.}~\bibnamefont {Paternostro}},\
  and\ \bibinfo {author} {\bibfnamefont {S.}~\bibnamefont {Campbell}},\
  }\bibfield  {title} {\bibinfo {title} {Quantum darwinism in a structured spin
  environment},\ }\href
  {https://doi.org/https://doi.org/10.1016/j.physleta.2021.127675} {\bibfield
  {journal} {\bibinfo  {journal} {Physics Letters A}\ }\textbf {\bibinfo
  {volume} {416}},\ \bibinfo {pages} {127675} (\bibinfo {year}
  {2021})}\BibitemShut {NoStop}%
\bibitem [{\citenamefont {Nahum}\ and\ \citenamefont
  {Wiese}(2023)}]{nahum2023renormalization}%
  \BibitemOpen
  \bibfield  {author} {\bibinfo {author} {\bibfnamefont {A.}~\bibnamefont
  {Nahum}}\ and\ \bibinfo {author} {\bibfnamefont {K.~J.}\ \bibnamefont
  {Wiese}},\ }\bibfield  {title} {\bibinfo {title} {Renormalization group for
  measurement and entanglement phase transitions},\ }\href
  {https://doi.org/10.1103/PhysRevB.108.104203} {\bibfield  {journal} {\bibinfo
   {journal} {Phys. Rev. B}\ }\textbf {\bibinfo {volume} {108}},\ \bibinfo
  {pages} {104203} (\bibinfo {year} {2023})}\BibitemShut {NoStop}%
\bibitem [{\citenamefont {Lieb}\ and\ \citenamefont
  {Robinson}(1972)}]{LRbound}%
  \BibitemOpen
  \bibfield  {author} {\bibinfo {author} {\bibfnamefont {E.~H.}\ \bibnamefont
  {Lieb}}\ and\ \bibinfo {author} {\bibfnamefont {D.~W.}\ \bibnamefont
  {Robinson}},\ }\bibfield  {title} {\bibinfo {title} {The finite group
  velocity of quantum spin systems},\ }\href
  {https://doi.org/10.1007/BF01645779} {\bibfield  {journal} {\bibinfo
  {journal} {Communications in Mathematical Physics}\ }\textbf {\bibinfo
  {volume} {28}},\ \bibinfo {pages} {251} (\bibinfo {year} {1972})}\BibitemShut
  {NoStop}%
\bibitem [{\citenamefont {Tran}\ \emph {et~al.}(2021)\citenamefont {Tran},
  \citenamefont {Guo}, \citenamefont {Baldwin}, \citenamefont {Ehrenberg},
  \citenamefont {Gorshkov},\ and\ \citenamefont {Lucas}}]{LRbound-LR}%
  \BibitemOpen
  \bibfield  {author} {\bibinfo {author} {\bibfnamefont {M.~C.}\ \bibnamefont
  {Tran}}, \bibinfo {author} {\bibfnamefont {A.~Y.}\ \bibnamefont {Guo}},
  \bibinfo {author} {\bibfnamefont {C.~L.}\ \bibnamefont {Baldwin}}, \bibinfo
  {author} {\bibfnamefont {A.}~\bibnamefont {Ehrenberg}}, \bibinfo {author}
  {\bibfnamefont {A.~V.}\ \bibnamefont {Gorshkov}},\ and\ \bibinfo {author}
  {\bibfnamefont {A.}~\bibnamefont {Lucas}},\ }\bibfield  {title} {\bibinfo
  {title} {Lieb-robinson light cone for power-law interactions},\ }\href
  {https://doi.org/10.1103/PhysRevLett.127.160401} {\bibfield  {journal}
  {\bibinfo  {journal} {Phys. Rev. Lett.}\ }\textbf {\bibinfo {volume} {127}},\
  \bibinfo {pages} {160401} (\bibinfo {year} {2021})}\BibitemShut {NoStop}%
\bibitem [{\citenamefont {Chen}\ \emph {et~al.}(2023)\citenamefont {Chen},
  \citenamefont {Lucas},\ and\ \citenamefont {Yin}}]{chen2023speed}%
  \BibitemOpen
  \bibfield  {author} {\bibinfo {author} {\bibfnamefont {C.-F.~A.}\
  \bibnamefont {Chen}}, \bibinfo {author} {\bibfnamefont {A.}~\bibnamefont
  {Lucas}},\ and\ \bibinfo {author} {\bibfnamefont {C.}~\bibnamefont {Yin}},\
  }\bibfield  {title} {\bibinfo {title} {Speed limits and locality in many-body
  quantum dynamics},\ }\href {https://doi.org/10.1088/1361-6633/acfaae}
  {\bibfield  {journal} {\bibinfo  {journal} {Reports on Progress in Physics}\
  }\textbf {\bibinfo {volume} {86}},\ \bibinfo {pages} {116001} (\bibinfo
  {year} {2023})}\BibitemShut {NoStop}%
\bibitem [{\citenamefont {Blume-Kohout}\ and\ \citenamefont
  {Zurek}(2008)}]{QD-QBM}%
  \BibitemOpen
  \bibfield  {author} {\bibinfo {author} {\bibfnamefont {R.}~\bibnamefont
  {Blume-Kohout}}\ and\ \bibinfo {author} {\bibfnamefont {W.~H.}\ \bibnamefont
  {Zurek}},\ }\bibfield  {title} {\bibinfo {title} {Quantum darwinism in
  quantum brownian motion},\ }\href
  {https://doi.org/10.1103/PhysRevLett.101.240405} {\bibfield  {journal}
  {\bibinfo  {journal} {Phys. Rev. Lett.}\ }\textbf {\bibinfo {volume} {101}},\
  \bibinfo {pages} {240405} (\bibinfo {year} {2008})}\BibitemShut {NoStop}%
\bibitem [{\citenamefont {Fert{\'e}}\ and\ \citenamefont
  {Cao}(2023)}]{ferte2023tree}%
  \BibitemOpen
  \bibfield  {author} {\bibinfo {author} {\bibfnamefont {B.}~\bibnamefont
  {Fert{\'e}}}\ and\ \bibinfo {author} {\bibfnamefont {X.}~\bibnamefont
  {Cao}},\ }\bibfield  {title} {\bibinfo {title} {Quantum darwinism-encoding
  transitions on expanding trees},\ }\bibfield  {journal} {\bibinfo  {journal}
  {Preprint}\ }\href {https://doi.org/10.48550/arXiv.2312.04284}
  {10.48550/arXiv.2312.04284} (\bibinfo {year} {2023})\BibitemShut {NoStop}%
\end{thebibliography}%

\newpage
\begin{widetext}

\section{Model of Quantum Darwinism-encoding transitions: solution details}

\subsection{Clifford-Operator approach to Quantum Darwinism: generalities}
In this section we explain in detail our general approach to Quantum Darwinism with Clifford unitary circuits. The results of this section do not depend on the tree structure of the models studied in the main text, and apply to a more general setup, depicted as follows:
\begin{equation}
	\includegraphics[scale=1,valign=c]{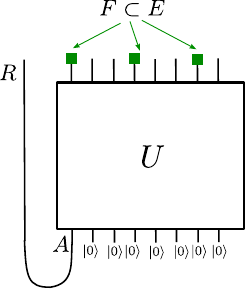}   
\end{equation}
where $U$ is any Clifford unitary acting on $N$ qubits. The $N$ input qubits are $A$, which is maximally entangled to the reference bit $R$, and the set $C$ of $(N-1)$ recruits that are initialized in a product state $| 0 \rangle^{\otimes (N-1)}$. For convenience, we shall also denote by $U$ the extended unitary $I_R \otimes U$ that acts trivially on the reference. We would like to understand the correlation between a subset $F$ of the set $E$ of output bits with $R$. 

We start by introducing some notations. Consider a set of qubits $X$ whose size is $|X|$. We shall denote by $\mathcal{P}_X$ the group generated by the Pauli operators acting $X$; the elements in $ \mathcal{P}_X$ are referred to as \textit{Pauli strings} on $X$. For example,  $\mathcal{P}_F$ is the set of all the (Pauli string) observable to which we have access.  The group structure of $ \mathcal{P}_X $ is defined by operator multiplication \textit{modulo a phase}. For example, we consider $ Z X = Y$ and $Z X Z = X$ to  be  valid identities, ignoring the factors $i$ and $-1$, respectively. (In general, the ignored phases are always a power of $i$.)  Then, $\mathcal{P}_X \simeq \mathbb{Z}_2^{2|X|}$ is isomorphic to the $2 |X|$-dimensional vector space over the finite field with two elements $\mathbb{Z}_2$. Hence, we should view $\mathcal{P}_X $ as a vector space.

A state $| \Psi \rangle$ on the qubit set $X$ is a \textit{stabilizer state} if there exists a subspace $ \mathcal{S}_\Psi \subset \mathcal{P}_X$ of dimension $|X|$, such that 
\begin{equation}
	\forall O \in \mathcal{S}_\Psi \,,\,   O | \Psi \rangle = \pm | \Psi \rangle  \,.
\end{equation} 
The space $ \mathcal{S}_\Psi$ is referred to as the \textit{stabilizer space}.
It follows that the initial state of the general setup, 
\begin{equation}
	{\Psi_0} = \frac1{\sqrt{2}} (|0_R 0_A \rangle + | 1_R 1_A \rangle)  \prod_{j\in C} |0_j \rangle
\end{equation}
is a stabilizer state (defined on the qubit set $RAC$). The stabilizer space is spanned by the following basis:
\begin{equation}\label{eq:S0}
	\mathcal{S}_{\Psi_0}  =  \mathrm{span} ( Z_R Z_A, X_R X_A, Z_1, Z_2, \dots, Z_{N-1}  )
\end{equation}
where $Z_j$ acts on the $j$-th recruit. In particular, as $R$ and $A$ form a maximally entangled pair, $Z_R $ and $Z_A$ are \textit{perfectly correlated}: $Z_R Z_A  |\Psi_0 \rangle =  | \Psi_0 \rangle $. The same can be said of $X_R$ and $X_A$. 

By definition, a Clifford unitary $U$ maps any Pauli operator to a Pauli string, and thus determines an isomorphism between the group of Pauli strings on the input and output qubits: 
\begin{equation} \label{eq:defU}
	\mathcal{U}: \mathcal{P}_{RE}  \to \mathcal{P}_{RAC}  \,,\,  O \mapsto  U^\dagger O U \,.
\end{equation}  
Note that the action of $\mathcal{U}$ is to ``pull back'' an operator on the output bits to one on the input ones. It follows that the final state $| \Psi_t \rangle := U \vert \Psi \rangle $ is also a stabilizer state, with the following stabilizer space:
\begin{equation} \label{eq:Spsit}
	\mathcal{S}_{ \Psi_t } =  	\mathcal{U}^{-1}  \mathcal{S}_{\Psi_0} \,.
\end{equation}  

Now, consider the sets of stabilizers supported on $F$ and on $RF$,  $\mathcal{S}_{ \Psi_t }  \cap \mathcal{P}_{ F } $ and $\mathcal{S}_{ \Psi_t }  \cap \mathcal{P}_{ F R } $, respectively. Both sets are $\mathbb{Z}_2$-vector spaces, and the former is contained in the latter. Their dimensions can differ by  at most $2$, since $\mathrm{dim}(\mathcal{P}_{ F R }) -  \mathrm{dim}(\mathcal{P}_{ F}) = 2$:
\begin{equation}
	\mathrm{dim}(\mathcal{S}_{ \Psi_t }  \cap \mathcal{P}_{FR}) -  \mathrm{dim}(\mathcal{S}_{ \Psi_t }  \cap \mathcal{P}_{F}) \in \{0, 1, 2 \} \,.
\end{equation} 
We now come to the first main result: \emph {the above dimension difference equal the mutual information between $F$ and $R$ in the unit of qubits}. Indeed, a classic result on stabilizer states is that the dimension of the stabilizer space $\mathcal{S}_X$ is related to the entanglement entropy (in unit of qubits, with respect to the state $\Psi_t$) of the subsystem $X$ as follows 
\begin{equation}
	H(X) = |X| - \mathrm{dim}(\mathcal{S}_{ \Psi_t }  \cap \mathcal{P}_{ X}) \,,
\end{equation}
where $|X|$ is the size of $X$. As a consequence, recalling that $H(R) = 1$, we have:
\begin{align}
	I(F,R) = H(F) + H(R) - H(FR) = & |F| - \mathrm{dim}(\mathcal{S}_{ \Psi_t }  \cap \mathcal{P}_{F}) + 1   -  |FR| + \mathrm{dim}(\mathcal{S}_{ \Psi_t }  \cap \mathcal{P}_{FR})  \nonumber \\ 
	= & \mathrm{dim}(\mathcal{S}_{ \Psi_t }  \cap \mathcal{P}_{FR}) -  \mathrm{dim}(\mathcal{S}_{ \Psi_t }  \cap \mathcal{P}_{F})  \,. \label{eq:Idimdiff}
\end{align} 

Furthermore,  the quotient space 
\begin{equation}\label{eq:defQ}
	\mathcal{Q} :=	\frac{\mathcal{S}_{ \Psi_t }  \cap \mathcal{P}_{FR}}{\mathcal{S}_{ \Psi_t }  \cap \mathcal{P}_{F}} \,,
\end{equation}
whose dimension equals $I(F,R) $, characterizes more explicitly the correlation between $F$ and $R$. Let us discuss case by case:
\begin{itemize}
	\item  When $ \mathrm{dim}(\mathcal{Q}) = 0$, there is no stabilizer supported on $FR$ that acts non-trivially on $R$. Thus, we cannot find any Pauli string on $F$ that is perfectly correlated with some non-identity Pauli on  $R$. This is expected from the zero mutual information. 
	\item  When $ \mathrm{dim}(\mathcal{Q}) = 1$, there is one stabilizer supported on $FR$ acting non-trivially on $R$. So it takes the form $P_R O_F$, where $P \in \{X, Y, Z\}$ and $O_F \in \mathcal{P}_F $. Therefore, $P_R O |\Psi_t \rangle = \pm  |\Psi_t \rangle $ and thus $O_F$ and $P_R$ are perfectly correlated~\footnote{$O$ cannot be identity; otherwise, we would have found a stabilizer supported on $R$, contradicting $H(R)=1$.}. Also, since $ \mathrm{dim}(\mathcal{Q}) = 1 $, two nonzero representatives of the quotient must differ by a stabilizer in $F$. Hence the Pauli $P_R$ is unique: there can be perfect correlation between some Pauli string in $F$ and one and only one Pauli on $R$. 
	\item Finally $ \mathrm{dim}(\mathcal{Q}) = 1$ means that we can two stabilizers $Z_R  O_F$ and $X_R O'_F$. They must commute, so $O_F$ and $O'_F$ must anti-commute, and can be viewed as the $Z$ and $X$ operators of some logical qubit. We have thus distilled a qubit from $F$ that is maximally entangled with $R$, as expected from $I(F,R) = 2$.
\end{itemize}

So far we have been focusing on the output qubits. The discussion is conceptually straightforward, but computationally inconvenient, since $\mathcal{S}_{\Psi_t}$ is cumbersome to describe directly. To overcome this, we shall use the isomorphism $\mathcal{U}$~\eqref{eq:defU} to pull back the quotient $\mathcal{Q}$ to the input bits. Using \eqref{eq:Spsit} and the fact that $U$ acts trivially on $R$, we have
\begin{equation}
	\mathcal{U}(\mathcal{Q}) =	\frac{ \mathcal{U} (\mathcal{S}_{ \Psi_t }  \cap \mathcal{P}_{FR}) }{ \mathcal{U} (\mathcal{S}_{ \Psi_t }  \cap \mathcal{P}_{F})} =  \frac{ \mathcal{S}_{ \Psi_0}  \cap \mathcal{U}(\mathcal{P}_{FR}) }{ \mathcal{S}_{ \Psi_0}  \cap \mathcal{U} (\mathcal{P}_{F}) }  =  \frac{ \mathcal{S}_{ \Psi_0}  \cap  ( \mathcal{U}(\mathcal{P}_{F}) +\mathcal{P}_{R})  }{ \mathcal{S}_{ \Psi_0}  \cap \mathcal{U} (\mathcal{P}_{F}) } \,.
\end{equation}
Now, recall that $ \mathcal{S}_{ \Psi_0} $ is explicitly known~\eqref{eq:S0}:  
\begin{equation}
	O \in \mathcal{S}_{ \Psi_0}  \text{ if and only if } O = O_R O_A \prod_{j \in C} Z_j^{e_j}   \label{eq:S0_form} 
\end{equation} 
for some $O \in \{I, X, Y, Z\}$ and $(e_j) \in \{0, 1\}^C$.  We now claim that the pull-backed quotient is isomorphic to the ``accessible subspace'' $\mathbf{s}$ described in the main text, and of which a formal definition is the following:
\begin{align} \label{eq:s-def}
	\mathbf{s} := \{ O_A \in \mathcal{P}_A | O_A = \mathrm{Tr}_C[\rho_C \, \mathcal{U}(O_F)]  \text{ for some }  O_F \in \mathcal{P}_F  \}   
\end{align}
Here $\rho_C = \prod_{j \in C }|0_j \rangle  \langle 0_j |$ is the initial density matrix of the recruits set $C$ and $\mathrm{Tr}_C$ is the partial trace on $C$. To show that claim, consider the linear map
\begin{align}
	\iota:  \mathcal{S}_{ \Psi_0}  \cap  ( \mathcal{U}(\mathcal{P}_{F}) +\mathcal{P}_{R})  \ni   O = O_R O_A \prod_{j \in C} Z_j^{e_j}   \mapsto O_A \in   \mathcal{P}_A \,.
\end{align} 
Now, $O \in \mathrm{ker}(\iota)$ if and only if $O_R = I$, which is equivalent to $ O \in \mathcal{S}_{ \Psi_0}  \cap  \mathcal{U}(\mathcal{P}_{F})$. Hence, $\iota$ induces an isomorphism from$  \mathcal{U}(\mathcal{Q})$ to its the image of $\iota$.  It remains to show that  $\mathrm{im}(\iota)  = \mathbf{s}$. Indeed, $O_A \in \mathrm{imag}(\iota) $ if and only if there is some $O_F \in \mathcal{P}_F$ such that $\mathcal{U}(O_F) = O_A \prod_{j \in C} Z_j^{e_j} $. And the last formula is equivalent to $ \mathrm{Tr}_C[\rho_C \mathcal{U}(O_F) ] = O_A.$  In summary, we have established the isomorphism
\begin{equation}
	\iota:  \mathcal{U}(\mathcal{Q})  \simeq \mathbf{s}  \,.
\end{equation}
In particular, $ \mathbf{s} $ is indeed a $\mathbb{Z}_2$-vector space. 

Combining  \eqref{eq:Idimdiff} and the above isomorphism, we conclude that the mutual information between $F$ and $R$ is equal to the dimension of $\mathbf{s}$
\begin{equation}
	I(F, R) = \mathrm{dim}(\mathbf{s}) \,.
\end{equation}
Similarly, combining the isomorphism $\iota  \mathcal{U}|_{\mathcal{Q}}$ and the discussion below \eqref{eq:defQ} about the information-theoretical meaning of $\mathcal{Q}$, we may see that $\mathbf{s}$ describes indeed the information on $R$ retrievable from $F$, as advocated in the main text. In particular, a non-identity Pauli $ O_A \in \mathbf{s} $ if and only if there is some Pauli-string observable in $F$ $O_F \in \mathcal{P}_F $ such that $P_R O_F | \Psi_t \rangle = \pm  | \Psi_t \rangle$, that is, $O_F$ is perfectly correlated with the same Pauli $O_R$ acting on the reference qubit.  

Before proceeding, we caution that, unlike $\mathcal{Q}$, $\mathbf{s}$ does not tells us which operators $O_F$ are perfectly correlated with $R$, but only their existence. Describing explicitly the possible $O_F$'s is in general a harder question. See however the end of Section~\ref{sec:recursion-analysis}.

\subsection{Backward recursion on a tree}
We showed above that the ``accessible subspace'' $\mathbf{s}$ [defined in \eqref{eq:s-def}] captures what $F$ knows about $R$. Now, we come to consider how to compute it (more precisely, its probability distribution) on a tree. For this it is more convenient to consider the isometry $V$ from the Hilbert space of $A$ to that of $E$ defined by the action of adjoining the recruit bits and applying $U$:
\begin{equation}
	V | \psi_A \rangle := U  \left( | \psi_A \rangle \otimes \prod_{j \in C} | 0_j \rangle \right) \,.
\end{equation}
In terms of $V$, the accessible subspace is given by the following:
\begin{equation}
	\mathbf{s} = 	\mathbf{s}(V, F) = \{  O_A \in \{I, X, Y, Z\} | O_A = V^\dagger O_F V \text{ for some } O_F \in \mathcal{P}_F \}. 
\end{equation}
In other words, $\mathbf{s}$ is the set of nonzero Pauli's (including identity) that can be obtained from pulling back a Pauli string on $F$ using the isometry $V$. 

Now, in our tree model, $\mathbf{s}$ can be computed recursively. The basic idea is best summarized in a picture: 
\begin{equation}
	\includegraphics[scale=1,valign=c]{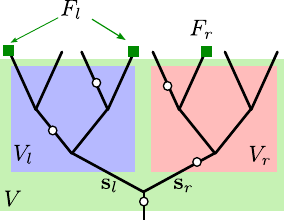}    \,.
\end{equation}
Indeed, the isometry $V$ can be built recursively:
\begin{equation}\label{eq:VVlVr}
	V   = ( V_l \otimes V_r )  \mathcal{B}  u   \,.
\end{equation}
Here, $V_l$ and $V_r$ are the isometries corresponding to the left and right subtrees, respectively.
$\mathcal{B}$
is the isometry corresponding to the branching node:
\begin{equation} \label{eq:isoB}
	\mathcal{B} := \sum_{i=0,1} |i \rangle | i \rangle \langle i |   \,.
\end{equation}
Finally $u$ is a one-qubit Clifford gate. It is equal to identity with probability $1-p$ and randomly chosen from the one-qubit Clifford group with probability $p$. 

It follows that $\mathbf{s}$ can be obtained rather simply from $\mathbf{s}_l = \mathbf{s} (V_l, F_l)$ and $\mathbf{s}_r = \mathbf{s} (V_r, F_r)$, where $F_{l}$ and $F_r$ are the set of accessible environment bits in the left and right subtree, respectively. More explicitly, we have
\begin{equation} \label{eq:recursive-relation-s}
	\mathbf{s} =  \sigma_u (\mathbf{B} (\mathbf{s}_l, \mathbf{s}_r ) ) . 
\end{equation}
Here, 
\begin{itemize}
	\item $	\sigma_u $ in \eqref{eq:recursive-relation-s} implements the pull back action of the one-body Clifford  gate:
	\begin{equation}
		\sigma_u (\mathbf{s}') :=  \{ u^\dagger O u : O \in \mathbf{s}' \} \,.
	\end{equation}
	If $u$ is identity, $\sigma_u (\mathbf{s}') =   \mathbf{s}'$ is the identity map. If $u$ is a random one-body Clifford, $\sigma_u$ acts as a random permutation on $\{ \mathbf{x},  \mathbf{y}, \mathbf{z}\}$ and leaves $ \mathbf{n}$ and $\mathbf{a}$ intact.
	\item $\mathbf{B}$ implements the pull-back action of the branching isometry \eqref{eq:isoB} on the accessible subsets:
	\begin{equation}
		\mathbf{B}:  (\mathbf{s}_1, \mathbf{s}_2)  \mapsto \{  \mathcal{B}^\dagger  (P_1 \otimes P_2)  \mathcal{B} | P_1 \in \mathbf{s}_{1} \,,\, P_2 \in \mathbf{s}_{2}  \} \cap \{I, Z, X, Y\}  \,.
	\end{equation}
	To compute it explicitly, we first calculate the map $ (P_1, P_2) \mapsto  \mathcal{B}^\dagger (P_1 \otimes P_2)  \mathcal{B} $. The result is the following table:
	\begin{equation}
		\begin{tabular}{|c|cccc|}
			\hline  & I & Z & X & Y  \\ \hline 
			I & I & Z & 0 & 0 \\ 
			Z & Z & I & 0 & 0 \\ 
			X & 0 & 0 & X & Y \\ 
			Y & 0 & 0 & Y & X \\ \hline
		\end{tabular} \,.
	\end{equation}
	It follows that the action of the map $\mathbf{B}$ on a pair of sets is given by another table:
	\begin{equation}\label{eq:tableau}
		\begin{tabular}{|c|ccccc|}
			\hline 
			$\mathbf{B}$ & $\mathbf{n}$ & $\mathbf{z}$  & $\mathbf{x}$ & $\mathbf{y}$ & $\mathbf{a}$ \\ \hline 
			$\mathbf{n}$ & $\mathbf{n}$ & $\mathbf{z}$ & $\mathbf{n}$ & $\mathbf{n}$ & $\mathbf{z}$ \\
			$\mathbf{z}$ & $\mathbf{z}$ & $\mathbf{z}$ & $\mathbf{z}$ & $\mathbf{z}$ & $\mathbf{z}$ \\
			$\mathbf{x}$ & $\mathbf{n}$ & $\mathbf{z}$ &  $\mathbf{x}$ & $\mathbf{y}$ & $\mathbf{a}$ \\
			$\mathbf{y}$ & $\mathbf{n}$ & $\mathbf{z}$ & $\mathbf{y}$ &  $\mathbf{x}$ & $\mathbf{a}$ \\
			$\mathbf{a}$ & $\mathbf{z}$ & $\mathbf{z}$ & $\mathbf{a}$ & $\mathbf{a}$  & $\mathbf{a}$ \\ \hline
		\end{tabular} \,.
	\end{equation}
\end{itemize}
The recursion relation above applies whenever the tree has at least one branching ($t > 0$). The initial case ($t=0$) is simple: $\mathbf{s} =\mathbf{a} $ if the only leaf of the tree is in $F$, and $\mathbf{s} =\mathbf{n} $ otherwise. 

For a given realization of $U$ and $F$, we can compute $\mathbf{s}$ by iterating the above recursion relation throughout the whole tree. We start by assigning $\mathbf{a}$ to the leaves in $F$ and $\mathbf{n}$ to those not in $F$, and then calculate the $\mathbf{s}$ associated with each subtree in a ``trickle-down'' manner,  from the leaves to the root. This is why such a method is known as a ``backward recursion''. Here is an illustration, with an arbitrarily chosen realization:
\begin{equation}
	\includegraphics[scale=1,valign=c]{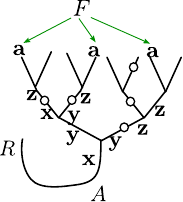} 
\end{equation}
Above, we omitted the $\mathbf{n}$'s and also wrote the intermediate outputs of the map $\mathbf{B}$. Also, recall that the absence of a $\circ$ means that the one-body unitary gate is identity, and its trivial action is not written. As a result, we obtain that $I(F, R) = 1$; more precisely, some Pauli string in $F$ is perfectly correlated with the $X$ operator in $R$.




\subsection{Backward recursion of the order parameter}
The above considerations apply to any fixed realization $(F,U)$. Now, since the random choices involved in $F$ and $U$ are all made locally (with respect to the tree geometry) and independently,  we can perform the average over all the realizations. As a result, we turn the above recursion relation into a recursion map that calculates the {probability distribution} of $\mathbf{s}$ of a $(t+1)$-generation tree, $\pi(t+1)$, in terms of $ \pi(t)$:
\begin{equation}\label{eq:f}
	{\pi}{(t+1)} = M({\pi}{(t)})  \,.
\end{equation}
Here, the recursion map $M$ is composition of two maps
\begin{equation}\label{eq:fPfy}
	M({\pi}) = M_u(M_{\mathcal{B}}({\pi})) 
\end{equation}
that we define now. $M_{\mathcal{B}}$ is a nonlinear map that implements the recursion relation between $\pi$'s at a branching, as follows:
\begin{align}\label{eq:fy}
	M_{\mathcal{B}}({\pi})_{\mathbf{s}}  = \sum_{\mathbf{s}_1,\mathbf{s}_2}   T^{\mathbf{s}}_{\mathbf{s1}\mathbf{s2}} \pi_{\mathbf{s1}} \pi_{\mathbf{s2}}  \,,\,   T^{\mathbf{s}}_{\mathbf{s1}\mathbf{s2}}  = \begin{dcases}
		1 & \mathbf{B}(\mathbf{s1}, \mathbf{s2}) = \mathbf{s} \\
		0 & \text{otherwise}.
	\end{dcases}
\end{align}
Explicitly, we have 
\begin{equation}
	\begin{bmatrix} M_{\mathcal{B}}({\pi})_{\mathbf{n}} \\ 
		M_{\mathcal{B}}({\pi})_{\mathbf{z}} \\ 
		M_{\mathcal{B}}({\pi})_{\mathbf{x}} \\
		M_{\mathcal{B}}({\pi})_{\mathbf{y}} \\
		M_{\mathcal{B}}({\pi})_{\mathbf{a}}
	\end{bmatrix} =  
	\begin{bmatrix}
		\pi_{\mathbf{n}}^2 + 2 \pi_{\mathbf{n}} (\pi_{\mathbf{x}} + \pi_{\mathbf{y}}) \\
		\pi_{\mathbf{z}}^2 + 2 \pi_{\mathbf{z}} (\pi_{\mathbf{n}} + \pi_{\mathbf{x}} + \pi_{\mathbf{y}} + \pi_{\mathbf{a}}) + 2 \pi_{\mathbf{n}} \pi_{\mathbf{a}} \\
		\pi_{\mathbf{x}}^2 + \pi_{\mathbf{y}}^2 \\
		2 \pi_{\mathbf{x}} \pi_{\mathbf{y}} \\
		\pi_{\mathbf{a}}^2 + 2 \pi_{\mathbf{a}} (\pi_{\mathbf{x}} + \pi_{\mathbf{y}}) 
	\end{bmatrix} \,. \nonumber
\end{equation}
$M_u$ is a linear map that accounts for the action of the random one-body Clifford (which exists with probability $1-p$):
\begin{align}\label{eq:fP}   
	M_u({y}) = (p P_3 + (1- p) I_5) {y}  \,,\,  P_3 = \frac13\begin{bmatrix} 3 & & & & \\
		&1&1&1& \\ 
		&1&1&1&\\ 
		&1&1&1&\\
		& & & & 3 \end{bmatrix}  \,.
\end{align}
The recursion relation above is supplemented by the initial condition:
\begin{equation} \label{eq:IC}
	{\pi}{(t=0)} = M_u (1-f, 0,0,0,f)^T = (1-f, 0,0,0,f)^T  \,.
\end{equation}
Indeed, a leaf belong to $F$ with probability $f$, in which case one has access to all of its Pauli's; otherwise, one has access to none. The map $M_u$ is applied since a random Clifford unitary can be applied to the output qubits; yet, this turns out to have no effect on the initial condition of the backward recursion. The recursion map $M$ defines a dynamical system on a space of dimension 4 defined by the sum rule
\begin{equation}
	\pi_{\mathbf{n}} +\pi_{\mathbf{z}} + \pi_{\mathbf{x}}+\pi_{\mathbf{y}} + \pi_{\mathbf{a}} =1  \,, \label{eq:sumrule}
\end{equation} 
since ${\pi}$ is a probability distribution. The phase diagram of the model is determined by the long-time limit of this dynamical system. (It is tempting call the dynamics generated by $M$ the ``renormalization group flow'' of the model. However we shall refrain from doing that since $\pi$'s are not really coupling constants.)

\subsection{Analysis of the recursion dynamics}\label{sec:recursion-analysis}
In this section we detail the analysis of the recursion dynamics. The results can be summarized as the flow diagrams shown in Figure~\ref{fig:flowM}. The method is a combination of analytics and numerics. We find all the fixed points of $M$ (which are the long time limits of $\pi$) analytically, as we detail below; we check the fixed points' stability numerically, by diagonalizing the linearization of $M$ around them. 

Let us start by discussing the symmetries. First, $M$ preserves the $\mathbb{Z}_2$ symmetry that swaps $\mathbf{n}$ and $\mathbf{a}$:  
\begin{equation}
	M(\tau({\pi})) = \tau(M({\pi})) \,,\, \text{where } \tau: (\pi_{\mathbf{n}},\pi_{\mathbf{z}}, \pi_{\mathbf{x}},\pi_{\mathbf{y}}, \pi_{\mathbf{a}}) \mapsto (\pi_{\mathbf{a}},\pi_{\mathbf{z}}, \pi_{\mathbf{x}},\pi_{\mathbf{y}}, \pi_{\mathbf{n}}) \,.
\end{equation}
The initial condition \eqref{eq:IC} breaks this symmetry unless $f = 1/2$; indeed, $\tau$ sends $f$ to $1-f$. Another (broken) symmetry of the model is that of $\mathcal{S}_3$, which permutes $\mathbf{x}, \mathbf{y}$ and $\mathbf{z}$ and leaves $\mathbf{n}$ and $\mathbf{a}$ intact. This symmetry is preserved by $M_u$ and the initial condition but broken by $M_{\mathcal{B}}$. 

Next, we consider invariant subspaces of the map $M$. First, 
\begin{equation}
	I_{xy} := \{  {\pi}: \pi_{\mathbf{x}} = \pi_{\mathbf{y}} \} \,,
\end{equation}
is invariant under $M$ and contains the initial conditions, so is where all the dynamics takes place (this can be seen as a ``relic'' of the broken $\mathcal{S}_3$ symmetry). Therefore, we shall restrict to $I_{xy}$ in what follows, and reduce the dimension of the dynamical system to 3. It is also useful to note another pair of invariant subspaces, 
\begin{equation}
	I_{+}  := \{  {\pi}: \pi_{\mathbf{a}} = 0 \}  \,,\,  I_{-}  :=  \{   {\pi}: \pi_{\mathbf{n}} = 0 \}  \,,
\end{equation}
which are related to each other by the $\mathbb{Z}_2$ symmetry.

The dynamics ${\pi} \mapsto M({\pi}) $ is structured by the fixed points of $M$. Indeed, numerically, we do not find any cycle with period $>1$ or other nontrivial asymptotic behaviors, and we do not expect these to occur on physical grounds. The fixed points $\pi^*$ satisfy the following independent equations [we used $\pi^*_{\mathbf{y}} = \pi^*_{\mathbf{x}}$):
\begin{align}
	& 3 (1-2 {\pi_{\mathbf{x}}^*}) {\pi_{\mathbf{x}}^*}=p \left(2 {\pi_{\mathbf{n}}^*} ({\pi_{\mathbf{z}}^*}+{\pi_{\mathbf{a}}^*})+{\pi_{\mathbf{z}}^*}^2+4 {\pi_{\mathbf{z}}^*} {\pi_{\mathbf{x}}^*}+2 {\pi_{\mathbf{z}}^*} {\pi_{\mathbf{a}}^*}-2 {\pi_{\mathbf{x}}^*}^2\right) \label{eq:fp-x}  \\ 
	& \pi^*_{\mathbf{a}} (\pi^*_{\mathbf{a}} + 4 \pi^*_{\mathbf{x}}) = \pi^*_{\mathbf{a}}  \,,\,  \pi^*_{\mathbf{n}} (\pi^*_{\mathbf{n}} + 4 \pi^*_{\mathbf{x}}) = \pi^*_{\mathbf{n}}  \,. \label{eq:fpan}
\end{align}
The last two equations are simple, and allow us to classify the fixed points according their membership with respect to  $I_+ $ and $I_-$. 

Outside $I_+ \cup I_- $, there is at most one fixed point. Indeed, such a fixed point must satisfy $ \pi^*_{\mathbf{a}} + 4 \pi^*_{\mathbf{x}} =  \pi^*_{\mathbf{n}} + 4^* \pi_{\mathbf{x}} = 1$, which, combined with the sum rule, gives $
\pi^*_{\mathbf{a}} =  \pi^*_{\mathbf{n}} = 1-4\pi^*_{\mathbf{x}} \,,\, \pi^*_{\mathbf{z}} = 6  \pi^*_{\mathbf{x}} - 1 \,.$
(So this fixed point preserves the $\mathbb{Z}_2$ symmetry.) Plugging these into \eqref{eq:fp-x}, we obtain a quadratic equation for $\pi^*_{\mathbf{x}}$
\begin{equation}
	(6 p-6) {\pi^*_{\mathbf{x}}}^2+(3-8 p) \pi^*_{\mathbf{x}} +p = 0 \,.
\end{equation}
whose positive solution is 
\begin{equation}\label{eq:fp1}
	\pi^*_{\mathbf{x}} = \frac{-\sqrt{40 p^2-24 p+9}+8 p-3}{12 (p-1)}  \,.
\end{equation}
This solution is physical (i.e. all components ${\pi}$ are positive) if and only if $p > 3/5$. When this is the case, the fixed point \eqref{eq:fp1} is always unstable with respect to a $\mathbb{Z}_2$-odd perturbation. Indeed, for $\epsilon$ small, we have
\begin{align*}
	M: ( \pi^*_{\mathbf{a}} - \epsilon, \dots ,  \pi^*_{\mathbf{n}} + \epsilon  ) \mapsto ( \pi^*_{\mathbf{a}} - \lambda \epsilon ,\dots,  \pi^*_{\mathbf{a}} + \lambda \epsilon) + \mathcal{O}(\epsilon^2) \,,
\end{align*}
with $\lambda = 1 + \pi^*_{\mathbf{a}} > 1$. Therefore, this fixed point does not characterize a stable phase. Instead, it is the long-time solution for $f = 1/2$ when $p > 3/5$, i.e., along the first-order transition line corresponding to the spontaneous breaking of the $\mathbb{Z}_2$ symmetry.

We now consider the fixed points in $I_+ \setminus I_-$, i.e., $\pi^*_{\mathbf{a}} = 0$ but $\pi^*_{\mathbf{n}} \ne 0$ (the fixed points in $I_- \setminus I_+$ can be then obtained by applying the $\mathbb{Z}_2$ symmetry). Then \eqref{eq:fpan} implies $\pi^*_\mathbf{n} = 1 - 4 \pi^*_\mathbf{x}$, and $ \pi^*_\mathbf{z} = 2 \pi^*_\mathbf{x} $. Plugging these into \eqref{eq:fp-x}, we find 
\begin{equation} \label{eq:fp2}
	\pi_\mathbf{x}^* \left(  \pi_\mathbf{x}^* - \frac{3-4p}{6(1-p)} \right) = 0 \,.
\end{equation}
This gives two fixed points: 
\begin{equation}
	\pi_* =  (1-u, u/2, u/4, u/4, 0) \,,\, u = \frac{6-8p}{3-3p}  \text{ or }  \pi_* = (1,0,0,0,0) \,.
\end{equation} 
When $p > 3/4$, only the second one is physical. We checked that it is also stable, and is the long-time limit for any initial condition with $f<1/2$: this is the encoding phase. When $p \in (3/5, 3/4)$, the fixed point $(1,0,0,0,0)$ becomes unstable, and $(1-u, u/2, u/4, u/4, 0) $ becomes physical, stable and the long-time limit for any initial condition with $f<1/2$: this is the mixed phase. [When $p < 3/5$, $1-u < 0$ so $(1-u, u/2, u/4, u/4, 0) $ is no longer physical.]

Finally, we look at the fixed points in $I_+ \cap I_-$, i.e., with $\pi^*_{\mathbf{a}} = \pi^*_{\mathbf{n}} = 0$. This means that $\pi^*_\mathbf{x} =  (1- \pi_\mathbf{z}^*)/2$. Plugging into \eqref{eq:fp-x}, we obtain 
\begin{equation}
	(3 p-3) {\pi_\mathbf{z}^*}^2 +(3-6 p) \pi_\mathbf{z}^* +p = 0 \,,
\end{equation}
whose positive solution is given by \eqref{eq:QD-sol} in the main text: 
\begin{equation}
	\pi_\mathbf{z}^* = \frac{\sqrt{24 p^2-24 p+9}+3-6 p}{6 (1-p)} \,.
\end{equation}
This fixed point is always physical, but only stable when $p > 3/5$, and is the long-time limit for any initial condition with $f \in (0,1)$: this is the QD phase. 

\begin{figure}
	\centering
	\includegraphics[width=1\textwidth]{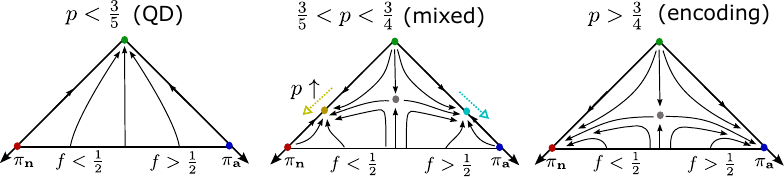}
	\caption{Flow diagram of the recursion map $M$ in QD, mixed and  encoding phases (from left to right), projected onto $\{ (\pi_{\mathbf{a}},\pi_{\mathbf{n}}): \pi_{\mathbf{a}} \ge 0, \pi_{\mathbf{n}} \ge 0, \pi_{\mathbf{a}} + \pi_{\mathbf{n}} \le 1 \}.$ The invariant subspaces $I_+$ and $I_-$ are projected on the axes. The $\mathbb{Z}_2$ symmetry acts by a left-right reflection. The initial conditions are at the bottom boundary.  The purple fixed point is \eqref{eq:fp1}. The yellow fixed point (only existing in the mixed phase) satisfies \eqref{eq:fp2}, and the cyan one is related by the  $\mathbf{Z}_2$ symmetry. The yellow and cyan arrows indicate how the these fix points move as $p$ increases. }
	\label{fig:flowM}
\end{figure}
We have thus found all the fixed points, and mapped out the phase diagram. In Figure~\ref{fig:flowM} we display the ``flow diagrams'' of the iteration dynamics in different phases. We remark that each phase is characterized by a unique stable fixed point, modulo the $\mathbb{Z}_2$ symmetry. Therefore, the long time limit of the recursion dynamics is essentially independent of the initial condition, except when we cross the first-order $\mathbb{Z}_2$-breaking transition $\{ f = 1/2, p > 3/5 \}$. In particular, the initial condition $(1-f, f, 0,0,0)$ results in the same long time limit as $(1-f, 0,0,0,f)$ as long as $f < 1/2$. Physically, this means that in the QD phase, we can retrieve a classical bit of the reference from a small fraction of the system $F$, even if we can only access the $Z$ operators on $F$. In the mixed phase, the same can be achieved in a nonzero fraction of the realizations.

\subsection{Joint distribution}
The analysis so far is about the ensemble of $(U, F)$, i.e., of one random circuit and one random subsystem.  To understand the nature of the mixed phase, it is useful to consider the joint ensemble of one random circuit and two random subsystems, $(U, F, G)$, such that $F \subset G$. They are constructed as follows: for any site $i$, we determine its membership with respect to $F$ and $G$ independently, with the following probabilities: 
\begin{align}
	\mathbb{P}( i \in F) = f \,,\, 
	\mathbb{P}( i \in G \setminus F) = g - f \,,\,  
	\mathbb{P}( i \notin G) = 1- g \,,
\end{align}
where $0 < f < g < 1$ are the relative size of $F$ and $G$, respectively. Note that if we forget about $F$ or $G$ from the joint ensemble, the remaining ensemble is the single subsystem ensemble we studied previously. 

For each realization $(U, F, G)$, we may construct two subgroups $\mathbf{s}$ and $\mathbf{t}$ (of accessible Pauli operators) with respect to $F$ and $G$, respectively. Then, the order parameter becomes the joint distribution $\Pi$ of $\mathbf{s}$ and $\mathbf{t}$, which is a $5\times 5$ matrix (rank 2 tensor). It is straightforward to derive the map $M_2$ that governs the backward recursion of $\Pi$, i.e., the analogue of $M$ \eqref{eq:f}. Like $M$, $M_2$ is a composition of two maps:
\begin{equation} \label{eq:M2}
	M_2(\Pi) = M_{2u}({M}_{2y}(\Pi)) \,.
\end{equation}
Here, $ M_{2\mathcal{B}}$ is the tensor square of $M_{\mathcal{B}}$~\eqref{eq:fy}:
\begin{equation}
	[M_{2\mathcal{B}}(\Pi)]_{\mathbf{s} \mathbf{t}} = 
	\sum_{\mathbf{s}_1 \mathbf{s}_2 \mathbf{t}_1  \mathbf{t}_2} T^{\mathbf{s}}_{\mathbf{s}_1 \mathbf{s}_2} T^{\mathbf{t}}_{\mathbf{t}_1 \mathbf{t}_2} \Pi_{\mathbf{s}_1 \mathbf{t}_1} \Pi_{\mathbf{s}_2 \mathbf{t}_2} \,.
\end{equation}
$M_{2u}$ is a linear map $\mathbb{R}^5 \otimes \mathbb{R}^5 \to \mathbb{R}^5 \otimes \mathbb{R}^5 $ defined as follows:
\begin{align}
	M_{2u} =  (1-p) \, I_5 \otimes  I_5  +   
	\frac{p}6 \sum_{\sigma \in \mathcal{S}_3} \diag(1, \sigma,1) \otimes \diag(1, \sigma,1) \,, \nonumber
\end{align} 
where $\sigma$ is summed over all permutation matrices of size $3$, and $\mathrm{diag}(1, \sigma, 1)$ denotes a block diagonal matrix with two blocks of size one (acting on $\mathbf{n}$ and $\mathbf{a}$) and a block of size $3$ (acting on $\mathbf{z},\mathbf{x},\mathbf{y}$). Note that $M_{2u}$ is \textit{not} the tensor square of $M_u$, since the \textit{same} random permutation acts on the two ``replicas'' with subsystems $F$ and $G$.  The initial condition for $\Pi$ satisfies
\begin{equation} \label{eq:M2-IC}
	\Pi{(t=0)}_{\mathbf{n}\mathbf{n}} = 1-g \,,\,   \Pi{(t=0)}_{\mathbf{n}\mathbf{a}} = g - f \,,\, 
	\Pi{(t=0)}_{\mathbf{a}\mathbf{a}} = f \,,
\end{equation}
and all the other components are zero.

\begin{figure}
	\centering
	\includegraphics[width=.85\textwidth]{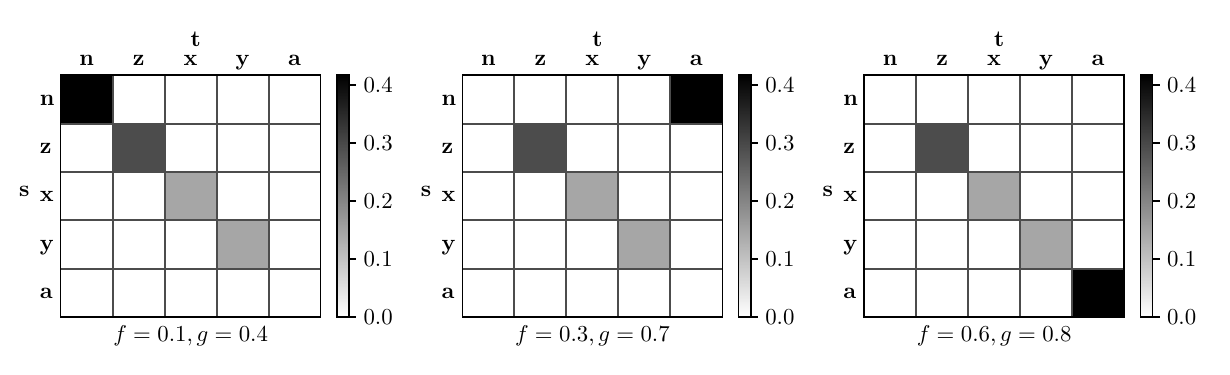}
	\caption{Examples of the joint distribution $ \Pi_{\mathbf{s} \mathbf{t}}(t\to\infty)$ in the mixed phase ($p = 0.32$), with different values of $f$ and $g$, obtained by solving numerically \eqref{eq:M2}-\eqref{eq:M2-IC} (one halts the iteration when a fixed point is reached up to numerical precision). One observes that $\mathbf{s}$ and $\mathbf{t}$ are always perfectly correlated, in the way described in text.  }
	\label{fig:joint}
\end{figure}
We studied numerically the recursion relation defined above in the mixed phase ($1/4< p < 2/5$). For all parameters $(p, f, g)$ tested, we found that $\Pi_{\mathbf{s} \mathbf{t}}$ has a nonzero limit only in one of the following situations (see Fig.~\ref{fig:joint} for an example):
\begin{itemize}
	\item $ \mathbf{s}= \mathbf{t} \in \{ \mathbf{x},  \mathbf{y},  \mathbf{z} \}$,
	\item $\mathbf{s} = \mathbf{t} = \mathbf{n} $, and $f < g < 1/2$; 
	\item $\mathbf{s} = \mathbf{t} = \mathbf{a} $, and $1/2 < f < g$; 
	\item $\mathbf{s} =  \mathbf{n}, \mathbf{t} = \mathbf{a} $, and $f < 1/2 < g$. 
\end{itemize}
The nonzero values of $\Pi_{\mathbf{s} \mathbf{t}}$ are completely determined by the fact that the marginal distributions must be equal to the one-point distribution ${\pi}$ studied above. Therefore, if we take a single mixed-phase realization in the $t\to\infty$ limit, and gradually increase the subsystem size fraction $f$, one and only one of the following scenario will take place:
\begin{itemize}
	\item We have a ``encoding-like'' realization in which $\mathbf{s}= \mathbf{n}$ as long as $f < 1/2$, and changes abruptly to $\mathbf{s}= \mathbf{a}$ for $f>1/2$. 
	\item We have a ``QD-Z'' realization in which $\mathbf{s}= \mathbf{z}$ for all $0<f<1$. 
	\item We have a ``QD-X'' or ``QD-Y'' realization, similarly defined.
\end{itemize}
In short, the mixed-phase ensemble is a macroscopic mixture of encoding-like realizations and QD-like ones.  ``Intermediate realizations'' cannot occur at the $t\to\infty$ limit, as the two-replica analysis would have suggested (see below). 

\subsection{Model with eavesdropping}
The above approach can be used to analyze the model with eavesdropping environment, upon making two modifications. First, the initial condition  should be modified to 
\begin{equation}
	\left[ \pi_{\mathbf{n}}, \pi_{\mathbf{z}},\pi_{\mathbf{x}}, \pi_{\mathbf{y}}, \pi_{\mathbf{a}}\right](t=0) = \left[1,0,0,0,0 \right] \,,
\end{equation}
since we have no access to the ``system'' output bits. Second, the recursion map $M$ is replaced by $   \tilde{M}$, defined as 
\begin{equation}
	\tilde{M}({\pi}) :=  (1-r) M_{p=1}({\pi}) + r M_e (M_{p=1}({\pi})) \,, 
\end{equation}
where 
\begin{equation}
	M_e({\pi})_{\mathbf{s}}  = \sum_{\mathbf{s}_1,\mathbf{s}_2}   T^{\mathbf{s}}_{\mathbf{s1}\mathbf{s2}} \pi_{\mathbf{s1}} \eta_{\mathbf{s2}} \,,\, \eta = [1-f, f, 0,0,0 ]\,,
\end{equation}
implements the eavesdropping [$T$ is defined in \eqref{eq:fy}]. Explicitly, 
\begin{equation}
	M_e({\pi}) =   f \, [0, 1,0,0,0]^T +  (1-f) \, [\pi_{\mathbf{n}} + \pi_{\mathbf{x}} + \pi_{\mathbf{y}}, \pi_{\mathbf{z}} + \pi_{\mathbf{a}}, 0,0,0 ]^T \,.
\end{equation}

To analyze the asymptotic behavior $\tilde{M}$, we first realize that the recursion dynamics is confined in the two-dimensional subspace 
\begin{equation}
	I_{xy} \cap I_+ = \left\{\pi_{\mathbf{a}} = 0, \pi_{\mathbf{x}} = \pi_{\mathbf{y}} \,,\,\pi_{\mathbf{z}} = 1 - \pi_{\mathbf{n}} - 2 \pi_{\mathbf{x}}  \right\} \,.
\end{equation} 
The fixed points are determined by two independent equations:
\begin{equation}
	\pi_\mathbf{x}^* = \frac13 (1-r) (1-\pi_\mathbf{n}^* (\pi_\mathbf{n}^* + 4 \pi_\mathbf{x}^*)) \,,\,   \pi_\mathbf{n}^* = \frac{r}{3} (1-f) (\pi_\mathbf{n}^* (\pi_\mathbf{n}^*+4 \pi_\mathbf{x}^*)+2) +\pi_\mathbf{n}^* (1-r) (\pi_\mathbf{n}^*+4 \pi_\mathbf{x}^*) \,.
\end{equation}
One may solve the first equation for $\pi_\mathbf{x}^*$ and plug the result into the second, which becomes a quadratic equation:
\begin{equation} \label{eq:MIPT-quadratic}
	(f r - 2 r + 1 ){\pi_\mathbf{n}^*}^2 + (4 r  -1 + 4 f r - 4 f r^2){\pi_\mathbf{n}^*} +2(f-1)r = 0 \,.
\end{equation}
We see immediately that $ \pi_\mathbf{n}^* = 0$ can be a solution if and only if $f=1$. 

When $f=1$, the other solution of \eqref{eq:MIPT-quadratic}
\begin{equation}\label{eq:pin-positive}
	\pi_\mathbf{n}^* = \frac{4r^2-8r+1}{1-r}   
\end{equation}
is negative when $r > r_c = \frac{1}{2} \left(2-\sqrt{3}\right)$. In this case, $ \pi_\mathbf{n}^* = 0$ is the only physical fixed point, and $\pi_{\mathbf{n}} \to 0$ in the $t\to\infty$ limit: this is the purified phase. When $r< r_c $, \eqref{eq:pin-positive} becomes a positive fixed point. It is also the only stable one, so $ \pi_{\mathbf{n}} \to  (4r^2-8r+1)/(1-r) > 0 $: we are in the ``volume law'' (or encoded) phase where there is nonzero probability that the environment fails to disentangle the reference from the system. 

When $f < 1$, \eqref{eq:MIPT-quadratic} always has one and only one positive root, and thus the $t\to\infty$ limit of $\pi_\mathbf{n} $ depends smoothly on $f$ and $r$. This means that there is no sharp MIPT-like transition if one can only access a fraction of the environment. When $f$ is close to $1$ and $r$ close to $r_c$, the singular part of $\pi_\mathbf{n}^*$ has the following single-parameter scaling behavior:
\begin{equation}
	\pi_\mathbf{n}^*  + 4 (r-r_c) \sim 4 |r-r_c|  \mathcal{F}\left( (1-f)^{\frac12} / |r-r_c| \right)  \,,\, \mathcal{F}(y) = \sqrt{y^2 /(4\sqrt{3}) + 1}  \,.
\end{equation}

\section{Details of the two-replica analysis}
We recall that, in general, the purity of a density matrix $\rho$ is equal to the expectation value of the partial swap operator that exchanges the two replicas,
\begin{equation}
	\Tr[\rho^2] = \Tr[(\rho \otimes \rho) \mathrm{SWAP}] \,,\, \mathrm{SWAP} | a \rangle | b \rangle := | b \rangle | a \rangle \,.
\end{equation}
To compute the annealed mutual information $I^{(2)}$, we should take $\rho$ to be the reduced density matrix of the subsystem $F$ or $FR$. Therefore, our strategy would be to consider the partial swap operator on $F$, evolve it backward with the circuit unitary $U$, and contract with the recruit bits. As a result, we obtain an operator $O$ acting on two copies of the qubit $A$, i.e., on the Hilbert space $(\mathbb{C}^2)^{\otimes 2}$. Since $A$ and $R$ formed a maximally entangled pair, we have 
\begin{equation} \label{eq:purtiyO}
	\Tr[ \rho_{F}^2 ] = \frac14  ( \sigma | O) \,,\, 
	\Tr[ \rho_{RF}^2 ] =  \frac14 ( \tau | O) \,.
\end{equation}
Here we introduced the operator inner product 
\begin{equation}
	(A|B) =  \Tr_{(\mathbb{C}^2)^{\otimes 2}}[ A B ] \,,
\end{equation}
and denoted by $\sigma$ and $\tau$ the identity and swap operator on $(\mathbb{C}^2)^{\otimes 2}$, respectively:
\begin{align}
	\langle j_2 j_1 |  \sigma |i_1 i_2 \rangle = \delta_{i_1 j_1} \delta_{i_2 j_2 }  \,,\ |\sigma) =  \includegraphics[scale=0.5,valign=c]{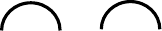}  \,,\, (\sigma | =  \includegraphics[scale=0.5,valign=c]{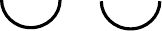}  \,,\,  \\  \langle j_2 j_1 |  \tau  |i_1 i_2 \rangle = \delta_{i_1 j_2} \delta_{i_2 j_1} \,,\,  |\tau ) = \includegraphics[scale=0.5,valign=c]{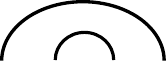} ,,\,  (\tau| = \includegraphics[scale=0.5,valign=c]{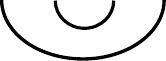} \,.
\end{align}
The inner product between these operator-states can be calculated by contracting the diagrams and associate a factor $q=2$ to each loop. Thus, the states $\sigma$ and $\tau$ are not orthonormal:
\begin{equation}  \label{eq:inner}
	( \sigma | \tau) = q = 2 \,,\,   ( \sigma | \sigma) =  ( \tau | \tau) = q^2 = 4 \,.
\end{equation}

Eq.~\eqref{eq:purtiyO} holds for any realization. Thus, to obtain the averaged purity, it suffices to calculate the average of $O$ over the subsystem $F$ and the random unitary $U$. On a tree, this can be done by a backward recursion. That is, we compute the average $  \overline{O{(t+1)}}  $ on a tree of $(t+1)$ generations, as a function of $  \overline{  O{(t)} } $. It is not hard to show that the recursion map is the following
\begin{equation}\label{eq:recursionO}
	| \overline{ O{(t+1)}} ) = ((1-p)+ p {L_u}  ) \,L_\mathcal{B} \, | \overline{ O{(t)}} )^{\otimes 2} \,.
\end{equation}
Here, the super-operator $L_\mathcal{B}$ evolves (backwards in time) an operator $O$ acting on 4 qubits (2 replicas per sub-tree) to an operator acting on 2 qubits via the isometry $\mathcal{B}$:
\begin{equation}\label{eq:defLY}
	L_\mathcal{B} O  := \mathcal{B}^{\otimes 2} O (\mathcal{B}^\dagger)^{\otimes 2} = \includegraphics[scale=.7,valign=c]{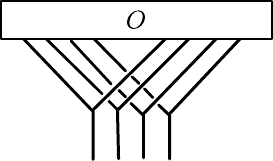}
\end{equation}
The super-operator ${L_u} $ implements the backward evolution by the random one-body Clifford unitary, replicated and averaged over the Clifford group (or equivalently over $U(2)$, since the Clifford group is a $2$-design): 
\begin{equation}\label{eq:defLu}
	{L_u} O := \left< (u \otimes u) O  (u^\dagger \otimes u^\dagger) \right>_u = \left< \includegraphics[scale=0.7,valign=c]{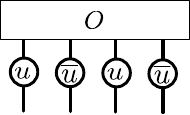}  \right>_u
\end{equation}
Equations \eqref{eq:recursionO} through \eqref{eq:defLu} define the backward recursion, which is supplemented by the initial condition:
\begin{equation}\label{eq:O-IC}
	\overline{  \vert  O{(t=0)}  )} = (1-f) \vert \sigma ) + f \vert \tau) \,.
\end{equation}

Now, the above recursion relation is significantly simplified by following fact: $\vert \overline{O{(t)}} )$ is always a linear combination of three operator-states:
\begin{equation}\label{eq:Ok-expansion}
	\vert \overline{O{(t)}} ) = w_{\sigma}{(t)} |\sigma ) + w_{\nu}{(t)} | \nu ) +  w_{\tau}{(t)}|\tau )  \,,
\end{equation}
where the new operator-state $\nu$ is defined as follows: 
\begin{equation}
	\langle  j_2 j_1 |  \nu | i_1 i_2 \rangle = \delta_{i_1 i_2 j_1 j_2} \,,\, | \nu ) = \includegraphics[scale=0.5,valign=c]{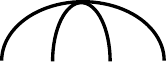} \,.
\end{equation}
Note that this expression is only valid in the spin-$Z$ basis. To see why \eqref{eq:Ok-expansion} must hold, and derive the recursion relation obeyed by the weights $w$, we shall calculate the action of $L_{\mathcal{B}}$ and $L_u$ on these states (and products thereof). Below, we report the results and briefly sketch their derivation.

The results for $L_{\mathcal{B}}$ can be summarized in a table
\begin{equation} \label{eq:Ly-table}
	\begin{tabular}{|c|ccc|}
		\hline
		$L_{\mathcal{B}}$  & $| \sigma )$  & $| \nu )$ &  $| \tau )$ \\ \hline
		$| \sigma )$ & $| \sigma )$  &  $| \nu )$ &  $| \nu )$   \\ 
		$| \nu )$ & $| \nu )$  &  $| \nu )$ &  $| \nu )$  \\
		$| \tau )$ & $| \nu )$  &  $| \nu )$ &  $| \tau )$ \\ \hline 
	\end{tabular}  
\end{equation}
That is, $L_{\mathcal{B}}$ maps all products between the three states to $| \nu )$  except $ | \sigma ) | \sigma ) \mapsto | \sigma)   $, $| \tau ) | \tau )  \mapsto | \tau) $. The latter cases follow from the fact that $\mathcal{B}$ is an isometry. The other cases can be seen diagrammatically as follows:
\begin{equation}
	\includegraphics[scale=0.8,valign=c]{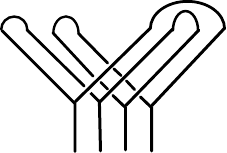} = \includegraphics[scale=0.8,valign=c]{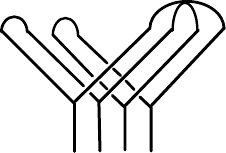} =
	\includegraphics[scale=0.8,valign=c]{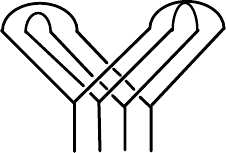}= \, \includegraphics[scale=0.7,valign=c]{nu1.pdf}  \,.
\end{equation}
Here, the first three diagrams represent $L_{\mathcal{B}} $ applied to $|\sigma) |\tau)$, $|\sigma) | \nu) $ and $|\tau) |\nu)$, respectively. In each case, the  diagram has a single connected component. This forces all the output indices to be equal, which is exactly what $|\nu)$ does.

For $L_u$,  we have 
\begin{equation} \label{eq:Lu-states}
	L_u | \sigma ) = | \sigma ) \,,\,   L_u | \tau )  = | \tau ) \,,\, 
	L_u | \nu ) = \frac1{3}  \left[ | \sigma ) +   | \tau ) \right] \,.
\end{equation}
[If $u$ is averaged over the unitary group $U(q)$, $1/3$ is replaced by $1/(q+1)$.] The first two equations of \eqref{eq:Lu-states} follow from unitary, whereas the last one can be derived using the Haar average formula (here $q=2$) 
\begin{align}
	\left< \vspace{.1cm}  \includegraphics[scale=.6,valign=c]{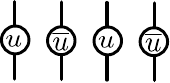} \right>_{U(q)} =& \frac1{q^2 - 1} \begin{matrix}  \includegraphics[scale=.5,valign=c]{sigma2.pdf} \\ 
		\includegraphics[scale=.5,valign=c]{sigma1.pdf} \end{matrix} + 
	\frac1{q^2 - 1} \begin{matrix} \includegraphics[scale=.5,valign=c]{tau2.pdf} \\ 
		\includegraphics[scale=.5,valign=c]{tau1.pdf} \end{matrix}  \nonumber \\
	&- \frac1{q(q^2 - 1)} \begin{matrix} \includegraphics[scale=.5,valign=c]{tau2.pdf}  \\ 
		\includegraphics[scale=.5,valign=c]{sigma1.pdf} \end{matrix} - 
	\frac1{q(q^2 - 1)} \begin{matrix} \includegraphics[scale=.5,valign=c]{sigma2.pdf} \\ 
		\includegraphics[scale=.5,valign=c]{tau1.pdf} \end{matrix} \,,  \nonumber
\end{align}
combined with $( \sigma | \nu) = ( \tau | \nu) = q$ (which can be checked diagrammatically).

Equations~\eqref{eq:Ly-table} and \eqref{eq:Lu-states} imply that $ \vert \overline{O{(t)}} ) $ is indeed a linear combination as in \eqref{eq:Ok-expansion}, and that the weights satisfy the recursion relation generated by the map:
\begin{equation}
	M_w: \begin{bmatrix} w_\sigma \\ w_{\nu} \\ w_\tau \end{bmatrix} \mapsto 
	\begin{bmatrix}
		1 & \frac{1-p}3 & 0 \\
		0 & p & 0 \\
		0  & \frac{1-p}3 & 1
	\end{bmatrix} \begin{bmatrix} w_\sigma^2 \\ w_{\nu}^2 + 2 (w_{\sigma} w_{\nu}+ w_{\sigma} w_{\tau} + w_{\nu} w_{\tau}) \\  w_\tau^2 \end{bmatrix} \,.
\end{equation}
By \eqref{eq:O-IC}, the initial condition is 
\begin{equation}
	(w_\sigma{(t=0)}, w_{\nu}{(t=0)}, w_\tau{(t=0)}) = (1-f, 0, f) \,.
\end{equation}

It remains to analyze the asymptotic behavior of the dynamical system generated by $M_w$. We observe that similarly to the recursion map $M$ above, $M_w$ preserves a $\mathbb{Z}_2$ symmetry, which acts by exchanging $\sigma$ and $\tau$; the new state $\nu$ is $\mathbb{Z}_2 $ even. However, there is a crucial difference: unlike $M$, $M_w$ does not preserve the sum $w_\sigma + w_\nu + w_\tau $. Indeed, $(w_\sigma, w_\nu, w_\tau)$ is not a probability distribution, but rather ``partition functions'' with different boundary conditions at the root of the tree. Yet, $w_\sigma, w_\nu, w_\tau$ are still non-negative. So we shall look for fixed points of $M_w$ up to a global factor. As a result of an elementary analysis, we find the following physically relevant fixed points: 
\begin{itemize}
	\item The ``encoding'' fixed points $(w_\sigma^*, w_\nu^*, w_\tau^*) \propto (1,0,0)$ and $(0,0,1)$. 
	\item The ``QD'' fixed point $(w_\sigma^*, w_\nu^*, w_\tau^*) \propto (u, 1-2u, u)$ where $u = u(p)$ is defined implicitly by inverting 
	\begin{equation}
		p = \frac{3 (1-u) u}{(u+1) \left(1-2 u^2\right)} \,, \label{eq:fpQD-renyi}
	\end{equation}
	and choosing the branch that increases from $u(p=0) = 0$ to $u(p=1) = 1/2$. 
	\item A pair of intermediate $\mathbb{Z}_2$ breaking fixed points $(w_\sigma^*, w_\nu^*, w_\tau^*) \propto (u_+, 1-u_+-u_-, u_-)$ where $u_{\pm}$ are the two roots of the equation 
	\begin{equation} \label{eq:intermediate-fixed}
		p u^2-(3-3 p) u+ 4 p-3 = 0\,.
	\end{equation}
	These fixed points are physical only when $3/4 \le p \le p_{l}$, where 
	\begin{equation}
		p_{l}= \frac{3}{7} \left(2 \sqrt{2}-1\right) \approx 0.783\dots  \,.
	\end{equation} 
	Both fixed points merge with the QD fixed point as $p \nearrow  p_{l} $ and  become complex as $p > p_{l}$.  They tend to the encoding fixed points as $p \searrow 3/4$, and become non-positive as $p < 3/4$. (See Figure~\ref{fig:flowRenyi}.) 
\end{itemize}
\begin{figure}
	\centering
	\includegraphics[width=\textwidth]{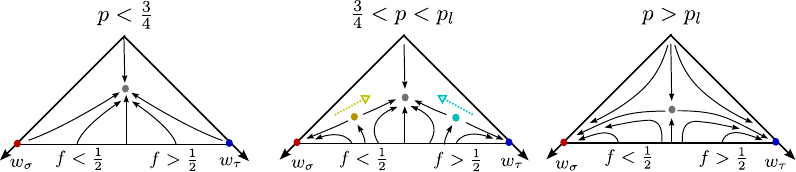}
	\caption{Flow diagram of the recursion map $M_w$ for $p<3/4$, $3/4 < p <p_{l}$ and  $p > p_{l} = \frac{3}{7} \left(2 \sqrt{2}-1\right) $ (from left to right). We consider $(w_\sigma, w_\nu, w_\tau)$ up to a global factor, and choose to normalize them so that they sum to $1$. The grey fixed point is \eqref{eq:fpQD-renyi}. The yellow and cyan fixed points satisfy \eqref{eq:intermediate-fixed}. When they exist, there are two threshold values of $f$ at which the asymptotic limit depends discontinuously on the initial condition. This gives rise to a $f$-dependent threshold $p_c(f)$, such that $p_c(1/2) = p_{l}$ and $p_c(0) = p_c(1) = 1/4$. The yellow and cyan arrows indicate how the intermediate fix points move as $p$ increases.}
	\label{fig:flowRenyi}
\end{figure}
Next we discuss the stability of the fixed points and the phase diagram, see Figure~\ref{fig:flowRenyi} for the flow diagram. The intermediate fixed points are always unstable. The encoding fixed points are stable when $p > 3/4$ and unstable when $p < 3/4$, and the QD fixed point is stable when $p < p_l$ and unstable when $p > p_l$. Therefore, 
\begin{itemize}
	\item When $p > p_l$, $w$ can only tend to the encoding fixed points, $\propto (1,0,0)$ if $f<1/2$ and $ \propto (0,0,1) $ if $f>1/2$. This means that $\overline{O} \propto \sigma $ or $\tau$, and $I^{(2)} = 0$ or $2$ in the two cases.
	\item  When $p < 3/4$, $w$ can only tend to the QD fixed point, $\propto (u, 1-2u, u)$. In this case $ \overline{O} \propto u (\sigma +\tau) + (1-2u) \nu$, and $I^{(2)} = 1$. 
	\item Finally, when $3/4  <p< p_l$, $w$ may tend to either an encoding or the QD fixed point, depending on the location of initial condition with respect to the stable manifold of the intermediate fixed points. Thus, there is a  $f$-dependent threshold $p_c(f) \in (3/4, p_l)$, such that $I^{(2)} \to 0$ or $2$ if $p > p_c(f)$ and $I^{(2)} \to 1 $ if $p < p_c(f)$. We have not found a closed-form expression of $p_c(f)$, and evaluated it numerically (see Figure \ref{fig:I2} of main text). It increases from $3/4$ to $p_{l}$ as $f$ increase from $0$ to $1/2$, and satisfies $p_c(f) = p_c(1-f)$. 
\end{itemize}
A notable consequence of our two-replica analysis is that the ``annealed $I$-$f$ curve'' when $p \in (3/4, p_l)$ is as follows:
\begin{equation} \label{eq:I2anomalous}
	I^{(2)}(R,F) \to \begin{dcases}
		0 & f < 1/2, p > p_c(f)  \\
		1 & p < p_c(f) \\ 
		2 & f > 1/2, p > p_c(f)
	\end{dcases}  \,.
\end{equation}
It differs from both the QD and encoding $I$-$f$ curves. Now, if one  assumed that $ I^{(2)}(R,F)$ equals $\overline{I(R,F)}$, the average of the true mutual information (which is a wrong assumption), then \eqref{eq:I2anomalous} would imply the existence (with nonzero probability as $t\to\infty$) of realizations whose $I$-$f$ curves differ from both QD and encoding curves. We have seen from the exact solution above that this is not the case! 
From the point of view of the flow diagram, the ``intermediate'' $I^{(2)}$-$f$ curve \eqref{eq:I2anomalous} occurs because of the existence of the unstable $\mathbb{Z}_2$-breaking fixed points. We have not found such fixed points from the replica-free analysis of any (Clifford) variants of our model that we studied (to be reported separately).

\end{widetext}

\end{document}